\documentclass[journal]{IEEEtran}
\usepackage{xcolor}
\usepackage{cite}
\usepackage{amsmath}
\usepackage{amsfonts}
\usepackage[short]{optidef}
\usepackage{subfig}
\usepackage[pdftex]{graphicx}
\usepackage{autobreak}
\usepackage{comment}
\usepackage{url}
\usepackage{booktabs}
\usepackage{amsmath}
\usepackage[acronyms,nonumberlist,nopostdot,nomain,nogroupskip,acronymlists={hidden}]{glossaries}
\newglossary[algh]{hidden}{acrh}{acnh}{Hidden Acronyms}
\newacronym{3gpp}{3GPP}{3rd Generation Partnership Project}
\newacronym{4g}{4G}{4th generation}
\newacronym{5g}{5G}{5th generation}
\newacronym{6g}{6G}{6th generation}
\newacronym{5gc}{5GC}{5G Core}
\newacronym{adc}{ADC}{Analog to Digital Converter}
\newacronym{aerpaw}{AERPAW}{Aerial Experimentation and Research Platform for Advanced Wireless}
\newacronym{ai}{AI}{Artificial Intelligence}
\newacronym{EM}{EM}{electromagnetic}
\newacronym{EMS}{EMS}{electromagnetic skin}
\newacronym{mmW}{mmW}{millimeter wave}
\newacronym{RoI}{RoI}{region of interest}
\newacronym{RIS}{RIS}{Reconfigurable Intelligent Surface}
\newacronym{RISs}{RISs}{Reconfigurable Intelligent Surfaces}
\newacronym{SNR}{SNR}{signal-to-noise ratio}
\newacronym{AF}{AF}{amplify-and-forward}
\newacronym{DF}{DF}{decode-and-forward}
\newacronym{3GPP}{3GPP}{3rd Generation Partnership Project}
\newacronym{RAN}{RAN}{Radio Access Network}
\newacronym{NCR}{NCR}{Network-Controlled Repeater}
\newacronym{NCRs}{NCRs}{Network-Controlled Repeaters}
\newacronym{IAB}{IAB}{Integated-Access-and-Backhauling}
\newacronym{SRE}{SRE}{Smart Radio Environment}
\newacronym{HSRE}{HSRE}{Heterogeneous SRE}
\newacronym{aimd}{AIMD}{Additive Increase Multiplicative Decrease}
\newacronym{am}{AM}{Acknowledged Mode}
\newacronym{amc}{AMC}{Adaptive Modulation and Coding}
\newacronym{amf}{AMF}{Access and Mobility Management Function}
\newacronym{aops}{AOPS}{Adaptive Order Prediction Scheduling}
\newacronym{api}{API}{Application Programming Interface}
\newacronym{apn}{APN}{Access Point Name}
\newacronym{ap}{AP}{Application Protocol}
\newacronym{aqm}{AQM}{Active Queue Management}
\newacronym{ar}{AR}{Augmented Reality}
\newacronym{ausf}{AUSF}{Authentication Server Function}
\newacronym{avc}{AVC}{Advanced Video Coding}
\newacronym{awgn}{AGWN}{Additive White Gaussian Noise}
\newacronym{balia}{BALIA}{Balanced Link Adaptation Algorithm}
\newacronym{bbu}{BBU}{Base Band Unit}
\newacronym{bdp}{BDP}{Bandwidth-Delay Product}
\newacronym{ber}{BER}{Bit Error Rate}
\newacronym{bf}{BF}{Beamforming}
\newacronym{bler}{BLER}{Block Error Rate}
\newacronym{brr}{BRR}{Bayesian Ridge Regressor}
\newacronym{BS}{BS}{Base Station}
\newacronym{bsr}{BSR}{Buffer Status Report}
\newacronym{bss}{BSS}{Business Support System}
\newacronym{ca}{CA}{Carrier Aggregation}
\newacronym{caas}{CaaS}{Connectivity-as-a-Service}
\newacronym{cb}{CB}{Code Block}
\newacronym{cc}{CC}{Congestion Control}
\newacronym{ccid}{CCID}{Congestion Control ID}
\newacronym{cco}{CC}{Carrier Component}
\newacronym{cdd}{CDD}{Cyclic Delay Diversity}
\newacronym{cdf}{CDF}{Cumulative Distribution Function}
\newacronym{cdn}{CDN}{Content Distribution Network}
\newacronym{cn}{CN}{Core Network}
\newacronym{codel}{CoDel}{Controlled Delay Management}
\newacronym{comac}{COMAC}{Converged Multi-Access and Core}
\newacronym{cord}{CORD}{Central Office Re-architected as a Datacenter}
\newacronym{cornet}{CORNET}{COgnitive Radio NETwork}
\newacronym{cosmos}{COSMOS}{Cloud Enhanced Open Software Defined Mobile Wireless Testbed for City-Scale Deployment}
\newacronym{cots}{COTS}{Commercial Off-the-Shelf}
\newacronym{cp}{CP}{Control Plane}
\newacronym{cyp}{CP}{Cyclic Prefix}
\newacronym{up}{UP}{User Plane}
\newacronym{cpu}{CPU}{Central Processing Unit}
\newacronym{cqi}{CQI}{Channel Quality Information}
\newacronym{cr}{CR}{Cognitive Radio}
\newacronym{cran}{C-RAN}{Cloud \gls{ran}}
\newacronym{crs}{CRS}{Cell Reference Signal}
\newacronym{csi}{CSI}{Channel State Information}
\newacronym{csirs}{CSI-RS}{Channel State Information - Reference Signal}
\newacronym{cu}{CU}{Central Unit}
\newacronym{d2tcp}{D$^2$TCP}{Deadline-aware Data center TCP}
\newacronym{d3}{D$^3$}{Deadline-Driven Delivery}
\newacronym{dac}{DAC}{Digital to Analog Converter}
\newacronym{dag}{DAG}{Directed Acyclic Graph}
\newacronym{das}{DAS}{Distributed Antenna System}
\newacronym{dash}{DASH}{Dynamic Adaptive Streaming over HTTP}
\newacronym{dc}{DC}{Dual Connectivity}
\newacronym{dccp}{DCCP}{Datagram Congestion Control Protocol}
\newacronym{dce}{DCE}{Direct Code Execution}
\newacronym{dci}{DCI}{Downlink Control Information}
\newacronym{dctcp}{DCTCP}{Data Center TCP}
\newacronym{dl}{DL}{Downlink}
\newacronym{dmr}{DMR}{Deadline Miss Ratio}
\newacronym{dmrs}{DMRS}{DeModulation Reference Signal}
\newacronym{drlcc}{DRL-CC}{Deep Reinforcement Learning Congestion Control}
\newacronym{drs}{DRS}{Discovery Reference Signal}
\newacronym{du}{DU}{Distributed Unit}
\newacronym{e2e}{E2E}{end-to-end}
\newacronym{ecaas}{ECaaS}{Edge-Cloud-as-a-Service}
\newacronym{ecn}{ECN}{Explicit Congestion Notification}
\newacronym{edf}{EDF}{Earliest Deadline First}
\newacronym{embb}{eMBB}{Enhanced Mobile Broadband}
\newacronym{empower}{EMPOWER}{EMpowering transatlantic PlatfOrms for advanced WirEless Research}
\newacronym{enb}{eNB}{evolved Node Base}
\newacronym{endc}{EN-DC}{E-UTRAN-\gls{nr} \gls{dc}}
\newacronym{epc}{EPC}{Evolved Packet Core}
\newacronym{eps}{EPS}{Evolved Packet System}
\newacronym{es}{ES}{Edge Server}
\newacronym{etsi}{ETSI}{European Telecommunications Standards Institute}
\newacronym[firstplural=Estimated Times of Arrival (ETAs)]{eta}{ETA}{Estimated Time of Arrival}
\newacronym{eutran}{E-UTRAN}{Evolved Universal Terrestrial Access Network}
\newacronym{faas}{FaaS}{Function-as-a-Service}
\newacronym{fapi}{FAPI}{Functional Application Platform Interface}
\newacronym{fdd}{FDD}{Frequency Division Duplexing}
\newacronym{fdm}{FDM}{Frequency Division Multiplexing}
\newacronym{fdma}{FDMA}{Frequency Division Multiple Access}
\newacronym{fed4fire}{FED4FIRE+}{Federation 4 Future Internet Research and Experimentation Plus}
\newacronym{fir}{FIR}{Finite Impulse Response}
\newacronym{fit}{FIT}{Future \acrlong{iot}}
\newacronym{fpga}{FPGA}{Field Programmable Gate Array}
\newacronym{fr2}{FR2}{Frequency Range 2}
\newacronym{fs}{FS}{Fast Switching}
\newacronym{fscc}{FSCC}{Flow Sharing Congestion Control}
\newacronym{ftp}{FTP}{File Transfer Protocol}
\newacronym{fw}{FW}{Flow Window}
\newacronym{ge}{GE}{Gaussian Elimination}
\newacronym{gnb}{gNB}{Next Generation Node Base}
\newacronym{gop}{GOP}{Group of Pictures}
\newacronym{gpr}{GPR}{Gaussian Process Regressor}
\newacronym{gpu}{GPU}{Graphics Processing Unit}
\newacronym{gtp}{GTP}{GPRS Tunneling Protocol}
\newacronym{gtpc}{GTP-C}{GPRS Tunnelling Protocol Control Plane}
\newacronym{gtpu}{GTP-U}{GPRS Tunnelling Protocol User Plane}
\newacronym{gtpv2c}{GTPv2-C}{\gls{gtp} v2 - Control}
\newacronym{gw}{GW}{Gateway}
\newacronym{harq}{HARQ}{Hybrid Automatic Repeat reQuest}
\newacronym{hetnet}{HetNet}{Heterogeneous Network}
\newacronym{hh}{HH}{Hard Handover}
\newacronym{hol}{HOL}{Head-of-Line}
\newacronym{hqf}{HQF}{Highest-quality-first}
\newacronym{hss}{HSS}{Home Subscription Server}
\newacronym{http}{HTTP}{HyperText Transfer Protocol}
\newacronym{ia}{IA}{Initial Access}
\newacronym{iab}{IAB}{Integrated Access and Backhaul}
\newacronym{ic}{IC}{Incident Command}
\newacronym{ietf}{IETF}{Internet Engineering Task Force}
\newacronym{imsi}{IMSI}{International Mobile Subscriber Identity}
\newacronym{imt}{IMT}{International Mobile Telecommunication}
\newacronym{iot}{IoT}{Internet of Things}
\newacronym{ip}{IP}{Internet Protocol}
\newacronym{itu}{ITU}{International Telecommunication Union}
\newacronym{kpi}{KPI}{Key Performance Indicator}
\newacronym{kpm}{KPM}{Key Performance Measurement}
\newacronym{kvm}{KVM}{Kernel-based Virtual Machine}
\newacronym{los}{LOS}{Line-of-Sight}
\newacronym{lsm}{LSM}{Link-to-System Mapping}
\newacronym{lstm}{LSTM}{Long Short Term Memory}
\newacronym{lte}{LTE}{Long Term Evolution}
\newacronym{lxc}{LXC}{Linux Container}
\newacronym{m2m}{M2M}{Machine to Machine}
\newacronym{mac}{MAC}{Medium Access Control}
\newacronym{manet}{MANET}{Mobile Ad Hoc Network}
\newacronym{mano}{MANO}{Management and Orchestration}
\newacronym{mc}{MC}{Multi-Connectivity}
\newacronym{mcc}{MCC}{Mobile Cloud Computing}
\newacronym{mchem}{MCHEM}{Massive Channel Emulator}
\newacronym{mcs}{MCS}{Modulation and Coding Scheme}
\newacronym{mec}{MEC}{Multi-access Edge Computing}
\newacronym{mec2}{MEC}{Mobile Edge Cloud}
\newacronym{mfc}{MFC}{Mobile Fog Computing}
\newacronym{mgen}{MGEN}{Multi-Generator}
\newacronym{mi}{MI}{Mutual Information}
\newacronym{mib}{MIB}{Master Information Block}
\newacronym{miesm}{MIESM}{Mutual Information Based Effective SINR}
\newacronym{mimo}{MIMO}{Multiple Input, Multiple Output}
\newacronym{ml}{ML}{Machine Learning}
\newacronym{mlr}{MLR}{Maximum-local-rate}
\newacronym[plural=\gls{mme}s,firstplural=Mobility Management Entities (MMEs)]{mme}{MME}{Mobility Management Entity}
\newacronym{mmtc}{mMTC}{Massive Machine-Type Communications}
\newacronym{mmwave}{mmWave}{millimeter wave}
\newacronym{mpdccp}{MP-DCCP}{Multipath Datagram Congestion Control Protocol}
\newacronym{mptcp}{MPTCP}{Multipath TCP}
\newacronym{mr}{MR}{Maximum Rate}
\newacronym{mrdc}{MR-DC}{Multi \gls{rat} \gls{dc}}
\newacronym{mse}{MSE}{Mean Square Error}
\newacronym{mss}{MSS}{Maximum Segment Size}
\newacronym{mt}{MT}{Mobile Termination}
\newacronym{mtd}{MTD}{Machine-Type Device}
\newacronym{mtu}{MTU}{Maximum Transmission Unit}
\newacronym{mumimo}{MU-MIMO}{Multi-user \gls{mimo}}
\newacronym{mvno}{MVNO}{Mobile Virtual Network Operator}
\newacronym{nalu}{NALU}{Network Abstraction Layer Unit}
\newacronym{nas}{NAS}{Non-Access Stratum}
\newacronym{nbiot}{NB-IoT}{Narrow Band IoT}
\newacronym{nfv}{NFV}{Network Function Virtualization}
\newacronym{nfvi}{NFVI}{Network Function Virtualization Infrastructure}
\newacronym{ngrg}{nGRG}{next Generation Research Group}
\newacronym{ni}{NI}{Network Interfaces}
\newacronym{nic}{NIC}{Network Interface Card}
\newacronym{nlos}{NLOS}{Non-Line-of-Sight}
\newacronym{now}{NOW}{Non Overlapping Window}
\newacronym{nsm}{NSM}{Network Service Mesh}
\newacronym{nr}{NR}{New Radio}
\newacronym{nrf}{NRF}{Network Repository Function}
\newacronym{nsa}{NSA}{Non Stand Alone}
\newacronym{nse}{NSE}{Network Slicing Engine}
\newacronym{nssf}{NSSF}{Network Slice Selection Function}
\newacronym{o2i}{O2I}{Outdoor to Indoor}
\newacronym{oai}{OAI}{OpenAirInterface}
\newacronym{oaicn}{OAI-CN}{\gls{oai} \acrlong{cn}}
\newacronym{oairan}{OAI-RAN}{\acrlong{oai} \acrlong{ran}}
\newacronym{oam}{OAM}{Operations, Administration and Maintenance}
\newacronym{ofdm}{OFDM}{Orthogonal Frequency Division Multiplexing}
\newacronym{olia}{OLIA}{Opportunistic Linked Increase Algorithm}
\newacronym{omec}{OMEC}{Open Mobile Evolved Core}
\newacronym{onap}{ONAP}{Open Network Automation Platform}
\newacronym{onf}{ONF}{Open Networking Foundation}
\newacronym{onos}{ONOS}{Open Networking Operating System}
\newacronym{oom}{OOM}{\gls{onap} Operations Manager}
\newacronym{opnfv}{OPNFV}{Open Platform for \gls{nfv}}
\newacronym{oran}{O-RAN}{Open Radio Access Network}
\newacronym{orbit}{ORBIT}{Open-Access Research Testbed for Next-Generation Wireless Networks}
\newacronym{os}{OS}{Operating System}
\newacronym{oss}{OSS}{Operations Support System}
\newacronym{otic}{OTIC}{Open Testing \& Integration Centre}
\newacronym{pa}{PA}{Position-aware}
\newacronym{pase}{PASE}{Prioritization, Arbitration, and Self-adjusting Endpoints}
\newacronym{pawr}{PAWR}{Platforms for Advanced Wireless Research}
\newacronym{pbch}{PBCH}{Physical Broadcast Channel}
\newacronym{pcef}{PCEF}{Policy and Charging Enforcement Function}
\newacronym{pcfich}{PCFICH}{Physical Control Format Indicator Channel}
\newacronym{pcrf}{PCRF}{Policy and Charging Rules Function}
\newacronym{pdcch}{PDCCH}{Physical Downlink Control Channel}
\newacronym{pdcp}{PDCP}{Packet Data Convergence Protocol}
\newacronym{pdsch}{PDSCH}{Physical Downlink Shared Channel}
\newacronym{pdu}{PDU}{Packet Data Unit}
\newacronym{pf}{PF}{Proportional Fair}
\newacronym{pgw}{PGW}{Packet Gateway}
\newacronym{phich}{PHICH}{Physical Hybrid ARQ Indicator Channel}
\newacronym{phy}{PHY}{Physical}
\newacronym{pmch}{PMCH}{Physical Multicast Channel}
\newacronym{pmi}{PMI}{Precoding Matrix Indicators}
\newacronym{powder}{POWDER}{Platform for Open Wireless Data-driven Experimental Research}
\newacronym{ppo}{PPO}{Proximal Policy Optimization}
\newacronym{ppp}{PPP}{Poisson Point Process}
\newacronym{prach}{PRACH}{Physical Random Access Channel}
\newacronym{prb}{PRB}{Physical Resource Block}
\newacronym{psnr}{PSNR}{Peak Signal to Noise Ratio}
\newacronym{pss}{PSS}{Primary Synchronization Signal}
\newacronym{pucch}{PUCCH}{Physical Uplink Control Channel}
\newacronym{pusch}{PUSCH}{Physical Uplink Shared Channel}
\newacronym{rar}{RAR}{Random Access Response}
\newacronym{qam}{QAM}{Quadrature Amplitude Modulation}
\newacronym{qci}{QCI}{\gls{qos} Class Identifier}
\newacronym{5qi}{5QI}{5G \gls{qos} Identifier}
\newacronym{qoe}{QoE}{Quality of Experience}
\newacronym{QoS}{QoS}{Quality of Service}
\newacronym{UE}{UE}{User Equipment}
\newacronym{UEs}{UEs}{User Equipments}
\newacronym{FoV}{FoV}{field of view}
\newacronym{UPA}{UPA}{uniform planar array}
\newacronym{quic}{QUIC}{Quick UDP Internet Connections}
\newacronym{rach}{RACH}{Random Access Channel}
\newacronym{ran}{RAN}{Radio Access Network}
\newacronym[firstplural=Radio Access Technologies (RATs)]{rat}{RAT}{Radio Access Technology}
\newacronym{rcn}{RCN}{Research Coordination Network}
\newacronym{STAR}{STAR RIS}{simultaneous transmitting and reflecting RIS}
\newacronym{3SNCR}{3SNCR}{trisectoral NCR}
\newacronym{rc}{RC}{RAN Control}
\newacronym{rec}{REC}{Radio Edge Cloud}
\newacronym{red}{RED}{Random Early Detection}
\newacronym{renew}{RENEW}{Reconfigurable Eco-system for Next-generation End-to-end Wireless}
\newacronym{rf}{RF}{Radio Frequency}
\newacronym{rfc}{RFC}{Request for Comments}
\newacronym{rfr}{RFR}{Random Forest Regressor}
\newacronym{ric}{RIC}{\gls{ran} Intelligent Controller}
\newacronym{rlc}{RLC}{Radio Link Control}
\newacronym{rlf}{RLF}{Radio Link Failure}
\newacronym{rlnc}{RLNC}{Random Linear Network Coding}
\newacronym{rmr}{RMR}{RIC Message Router}
\newacronym{rmse}{RMSE}{Root Mean Squared Error}
\newacronym{rnis}{RNIS}{Radio Network Information Service}
\newacronym{rr}{RR}{Round Robin}
\newacronym{rrc}{RRC}{Radio Resource Control}
\newacronym{rrm}{RRM}{Radio Resource Management}
\newacronym{rru}{RRU}{Remote Radio Unit}
\newacronym{rs}{RS}{Remote Server}
\newacronym{rsrp}{RSRP}{Reference Signal Received Power}
\newacronym{rsrq}{RSRQ}{Reference Signal Received Quality}
\newacronym{rss}{RSS}{Received Signal Strength}
\newacronym{rssi}{RSSI}{Received Signal Strength Indicator}
\newacronym{rtt}{RTT}{Round Trip Time}
\newacronym{ru}{RU}{Radio Unit}
\newacronym{rw}{RW}{Receive Window}
\newacronym{rx}{RX}{Receiver}
\newacronym{s1ap}{S1AP}{S1 Application Protocol}
\newacronym{sa}{SA}{standalone}
\newacronym{sack}{SACK}{Selective Acknowledgment}
\newacronym{sap}{SAP}{Service Access Point}
\newacronym{sc2}{SC2}{Spectrum Collaboration Challenge}
\newacronym{scef}{SCEF}{Service Capability Exposure Function}
\newacronym{sch}{SCH}{Secondary Cell Handover}
\newacronym{scoot}{SCOOT}{Split Cycle Offset Optimization Technique}
\newacronym{sctp}{SCTP}{Stream Control Transmission Protocol}
\newacronym{sdap}{SDAP}{Service Data Adaptation Protocol}
\newacronym{sdk}{SDK}{Software Development Kit}
\newacronym{sdm}{SDM}{Space Division Multiplexing}
\newacronym{sdma}{SDMA}{Spatial Division Multiple Access}
\newacronym{sdn}{SDN}{Software-defined Networking}
\newacronym{sdr}{SDR}{Software-defined Radio}
\newacronym{seba}{SEBA}{SDN-Enabled Broadband Access}
\newacronym{sgsn}{SGSN}{Serving GPRS Support Node}
\newacronym{sgw}{SGW}{Service Gateway}
\newacronym{si}{SI}{Study Item}
\newacronym{sib}{SIB}{Secondary Information Block}
\newacronym{sinr}{SINR}{Signal to Interference plus Noise Ratio}
\newacronym{sip}{SIP}{Session Initiation Protocol}
\newacronym{siso}{SISO}{Single Input, Single Output}
\newacronym{sla}{SLA}{Service Level Agreement}
\newacronym{sm}{SM}{Service Model}
\newacronym{smf}{SMF}{Session Management Function}
\newacronym{smo}{SMO}{Service Management and Orchestration}
\newacronym{sms}{SMS}{Short Message Service}
\newacronym{smsgmsc}{SMS-GMSC}{\gls{sms}-Gateway}
\newacronym{snr}{SNR}{Signal-to-Noise-Ratio}
\newacronym{son}{SON}{Self-Organizing Network}
\newacronym{sptcp}{SPTCP}{Single Path TCP}
\newacronym{srb}{SRB}{Service Radio Bearer}
\newacronym{srn}{SRN}{Standard Radio Node}
\newacronym{srs}{SRS}{Sounding Reference Signal}
\newacronym{zc}{ZC}{Zadoff-Chu}
\newacronym{ta}{TA}{Timing Advance}
\newacronym{ss}{SS}{Synchronization Signal}
\newacronym{sss}{SSS}{Secondary Synchronization Signal}
\newacronym{st}{ST}{Spanning Tree}
\newacronym{svc}{SVC}{Scalable Video Coding}
\newacronym{tb}{TB}{Transport Block}
\newacronym{tcp}{TCP}{Transmission Control Protocol}
\newacronym{tdd}{TDD}{Time Division Duplexing}
\newacronym{tdm}{TDM}{Time Division Multiplexing}
\newacronym{tdma}{TDMA}{Time Division Multiple Access}
\newacronym{tfl}{TfL}{Transport for London}
\newacronym{tfrc}{TFRC}{TCP-Friendly Rate Control}
\newacronym{tft}{TFT}{Traffic Flow Template}
\newacronym{tgen}{TGEN}{Traffic Generator}
\newacronym{tip}{TIP}{Telecom Infra Project}
\newacronym{tm}{TM}{Transparent Mode}
\newacronym{to}{TO}{Telco Operator}
\newacronym{tr}{TR}{Technical Report}
\newacronym{trp}{TRP}{Transmitter Receiver Pair}
\newacronym{ts}{TS}{Technical Specification}
\newacronym{tti}{TTI}{Transmission Time Interval}
\newacronym{ttt}{TTT}{Time-to-Trigger}
\newacronym{tx}{TX}{Transmitter}
\newacronym{uas}{UAS}{Unmanned Aerial System}
\newacronym{uav}{UAV}{Unmanned Aerial Vehicle}
\newacronym{udm}{UDM}{Unified Data Management}
\newacronym{udp}{UDP}{User Datagram Protocol}
\newacronym{udr}{UDR}{Unified Data Repository}
\newacronym{ue}{UE}{User Equipment}
\newacronym{uhd}{UHD}{\gls{usrp} Hardware Driver}
\newacronym{ul}{UL}{Uplink}
\newacronym{um}{UM}{Unacknowledged Mode}
\newacronym{uml}{UML}{Unified Modeling Language}
\newacronym{upa}{UPA}{Uniform Planar Array}
\newacronym{upf}{UPF}{User Plane Function}
\newacronym{urllc}{URLLC}{Ultra Reliable and Low Latency Communications}
\newacronym{usa}{U.S.}{United States}
\newacronym{usim}{USIM}{Universal Subscriber Identity Module}
\newacronym{usrp}{USRP}{Universal Software Radio Peripheral}
\newacronym{utc}{UTC}{Urban Traffic Control}
\newacronym{vim}{VIM}{Virtualization Infrastructure Manager}
\newacronym{vm}{VM}{Virtual Machine}
\newacronym{vnf}{VNF}{Virtual Network Function}
\newacronym{volte}{VoLTE}{Voice over \gls{lte}}
\newacronym{voltha}{VOLTHA}{Virtual OLT HArdware Abstraction}
\newacronym{vr}{VR}{Virtual Reality}
\newacronym{vran}{vRAN}{Virtualized \gls{ran}}
\newacronym{vss}{VSS}{Video Streaming Server}
\newacronym{wbf}{WBF}{Wired Bias Function}
\newacronym{wf}{WF}{Waterfilling}
\newacronym{wg}{WG}{Working Group}
\newacronym{wlan}{WLAN}{Wireless Local Area Network}
\newacronym{osm}{OSM}{Open Source Management and Orchestration}
\newacronym{pnf}{PNF}{Physical Network Function}
\newacronym{drl}{DRL}{Deep Reinforcement Learning}
\newacronym{mtc}{MTC}{Machine-type Communications}
\newacronym{osc}{OSC}{O-RAN Software Community}
\newacronym{mns}{MnS}{Management Services}
\newacronym{ves}{VES}{\gls{vnf} Event Stream}
\newacronym{ei}{EI}{Enrichment Information}
\newacronym{fh}{FH}{Fronthaul}
\newacronym{fft}{FFT}{Fast Fourier Transform}
\newacronym{laa}{LAA}{Licensed-Assisted Access}
\newacronym{plfs}{PLFS}{Physical Layer Frequency Signals}
\newacronym{ptp}{PTP}{Precision Time Protocol}
\newacronym{asic}{ASIC}{Application-specific Integrated Circuit}
\newacronym{aal}{AAL}{Acceleration Abstraction Layer}
\newacronym{fec}{FEC}{Forward Error Correction}
\newacronym{sdl}{SDL}{Shared Data Layer}
\newacronym{nib}{NIB}{Network Information Base}
\newacronym{rnib}{R-NIB}{RAN \gls{nib}}
\newacronym{fcaps}{FCAPS}{Fault, Configuration, Accounting, Performance, Security}
\newacronym{ie}{IE}{Information Element}
\newacronym{fg}{FG}{Focus Group}
\newacronym{osfg}{OSFG}{Open Source Focus Group}
\newacronym{sdfg}{SDFG}{Standard Development Focus Group}
\newacronym{tifg}{TIFG}{Test \& Integration Focus Group}
\newacronym{sfg}{SFG}{Security Focus Group}
\newacronym{swg}{SWG}{Security Work Group}
\newacronym{e2sm}{E2SM}{E2 Service Model}
\newacronym{tsc}{TSC}{Technical Steering Committee}
\newacronym{sdo}{SDO}{Standard-Development Organization}
\newacronym{sql}{SQL}{Structured Query Language}
\newacronym{ssh}{SSH}{Secure Shell}
\newacronym{tls}{TLS}{Transport Layer Security}
\newacronym{netconf}{NETCONF}{Network Configuration Protocol}
\newacronym{dtls}{DTLS}{Datagram Transport Layer Security}
\newacronym{cmp}{CMP}{Certificate Management Protocol}
\newacronym{ccc}{CCC}{Cell Configuration and Control}
\newacronym{dsp}{DSP}{Digital Signal Processing}
\newacronym{opex}{OPEX}{Operational Expenses}
\newacronym{cbrs}{CBRS}{Citizen Broadband Radio Service}
\newacronym{ntn}{NTN}{Non-terrestrial Network}
\newacronym{gbr}{GBR}{Guaranteed Bitrate}
\newacronym{sps}{SPS}{Semi-Persistent Scheduling}
\newacronym{tbs}{TBS}{Transport Block Size}
\newacronym{gnss}{GNSS}{Global Navigation Satellite System}
\newacronym{tof}{ToF}{Time of Flight}
\newacronym{rtof}{RToF}{Return Time of Flight}
\newacronym{rsig}{RS}{Reference Signal}
\newacronym{nrtric}{near-RT RIC}{near-Real Time Ran Intelligent Controller}
\newacronym{nonrtric}{non-RT RIC}{non-Real Time Ran Intelligent Controller}
\newacronym{aoa}{AoA}{Angle of Arrival}
\newacronym{tdoa}{TDoA}{Time Difference of Arrival}
\newacronym{rtoa}{RToA}{Return Time of Arrival}
\newacronym{ecdf}{ECDF}{Empirical Cumulative Distribution Function}
\newacronym{ris}{RIS}{Reconfigurable Intelligent Surface}
\newacronym{srd}{SRD}{Smart Radio Device}
\newacronym{gfbr}{GFBR}{Guaranteed Flow Bit Rate}
\newacronym{rg}{RG}{Resource Grid}
\newacronym{rb}{RB}{Resource Block}
\newacronym{re}{RE}{Resource Element}
\newacronym{rfra}{RF}{Radio Frame}
\newacronym{scs}{SCS}{Subcarrier Spacing}
\newacronym{ec}{EC}{Edge Computing}
\newacronym{af}{AF}{Amplify-and-Forward}
\newacronym{ncr}{NCR}{Network-Controlled Repeater}
\newacronym{tp}{TP}{Test Point}
\newacronym{cs}{CS}{Candidate Site}
\newacronym{src}{SRC}{Smart Radio Connection}
\newacronym{milp}{MILP}{Mixed Integer-Linear Programming}

\newacronym{FCMC}{FCMC}{full coverage minimum cost}

\newacronym{MBCC}{MBCC}{maximum budget-constrained coverage}

\newacronym{PDF}{PDF}{probability density function}
\begin{document}
\title{Optimal Planning for \\ Heterogeneous Smart Radio Environments}

\author{Reza~Agahzadeh~Ayoubi$^{*}$~\IEEEmembership{Member,~IEEE,}
        Eugenio~Moro$^{*}$,~\IEEEmembership{Member,~IEEE,}   \\
        Marouan~Mizmizi,~\IEEEmembership{Member,~IEEE,}
        Dario~Tagliaferri,~\IEEEmembership{Member,~IEEE,}\\
        Ilario~Filippini,~\IEEEmembership{Senior~Member,~IEEE,}
        Umberto~Spagnolini,~\IEEEmembership{Senior~Member,~IEEE}
\thanks{This work was partially supported by the European Union - Next Generation EU under the Italian National Recovery and Resilience Plan (NRRP), Mission 4, Component 2, Investment 1.3, CUP D43C22003080001, partnership on “Telecommunications of the Future” (PE00000001 - program “RESTART”)}
\thanks{The authors are with the Department of electronics, Information and Bioengineering, Politecnico di Milano, 20133, Milano, Italy}
\thanks{$^{*}$ These authors contributed equally to this research.}}

\maketitle
\begin{abstract}
\gls{SRE} is a central  paradigms in 6G and beyond, where integrating \gls{SRE} components into the network planning process enables optimized performance for high-frequency \gls{RAN}. This paper presents a comprehensive planning framework utilizing realistic urban scenarios and precise channel models to analyze diverse \gls{SRE} components, including \gls{RIS}, \gls{NCR}, and advanced technologies like \gls{STAR} and \gls{3SNCR}. We propose two optimization methods—\gls{FCMC} and \gls{MBCC}—that address key cost and coverage objectives by considering both physical characteristics and scalable costs of each component, influenced by factors such as \gls{NCR} amplification gain and \gls{RIS} dimensions. Extensive numerical results demonstrate the significant impact of these models in enhancing network planning efficiency for high-density urban environments.
\end{abstract}

\glsresetall
\section{Introduction}
The ever-growing demand for data envisioned for 6G as well as the evolution of communication technologies necessitates the exploration of new frequency bands, in the centimeter-wave ($6-24$ GHz), a.k.a. mid-band, and  \gls{mmW} spectrum portions ($24-100$ GHz). A notable factor of these bands is their harsh propagation characteristics and the high sensitivity to link blockage, which limit the network coverage. Therefore, looking beyond 5G, the future of \gls{RAN} architectures hinges on two pivotal advancements. On one side, \gls{IAB} allows to miigate these issues by network densification providing in-band backhauling, thereby enhancing network efficiency and reducing deployment costs \cite{IAB1}. On the other hand, from the physical layer perspective, the concept of a \gls{SRE} is emerging \cite{SRE_2,Renzo2020}. Here, the environment is no longer a static input but an adaptable and dynamic entity, allowing for dynamic manipulation to improve communication performance. \gls{SRE} includes two distinct devices, namely \gls{RIS}s and \gls{NCR}s \cite{SRE,makki_planning}. \gls{SRE} and \gls{IAB} are no muually exclusive; indeed \glspl{RIS} and \glspl{NCR} can be deployed o improve the communication performance of any generic \gls{RAN}, including those based on \gls{IAB} architecures.

\gls{RIS}s are quasi-passive devices that allow controlling the reflection/transmission direction of an impinging electromagnetic (EM) wave by tuning the physical properties of the reflection coefficient of the device itself \cite{susceptiblity}. At microwaves, \gls{RIS}s are usually manufactured as 2D arrays of unit cells (meta-atoms), whose reflection/transmission coefficient can be dynamically tuned to obtained the desired functionality, following the generalized Snell's law of reflection/refraction \cite{GSnell}. When \gls{RIS}s operate in both reflection and transmission mode, we properly refer to\gls{STAR} \cite{STAR_Schober}.

Differently, \gls{NCR}s are active devices that extend traditional amplify and forward relays with proper beamforming capabilities and time-division
duplex operation \cite{SRE}. \gls{NCR}s are introduced in recent release 18 of the \gls{3GPP} specifications \cite{3gpp_18_ncr} and are typically made by two analog antenna arrays (panels), one oriented towards the base station (BS) and the other to serve the users in a pre-defined area. Recently, a \gls{3SNCR} made by three panels has been introduced to extend the coverage of the conventional two-panel one~\cite{3SectSR}. 

The integration of \gls{HSRE} into future communication networks presents an opportunity to significantly enhance both coverage and overall quality of service (QoS). However, to fully leverage \gls{HSRE} capabilities, it is crucial to adapt the network planning process, which is the focus of this work. In the following, we review the literature on \gls{SRE} and \gls{HSRE}, with a specific emphasis on network planning.

\subsection{Literature on Network Planning within the \gls{HSRE}}\label{sec:related_work_SRE}
The literature on \gls{HSRE} comprises works on both \gls{RIS}s and \gls{NCR}s. For a single transmitting (Tx) -- receiving (Rx) pair, the optimal placement of the \gls{RIS} is explored in works such as \cite{RIS_delpoymen_Pos, RIS_delpoymen_Pos2}. For simplified scenarios, where blockage is not considered, it is preferable to place the \gls{RIS} either close to the Tx or Rx to reduce the overall path-loss of the cascaded Tx-\gls{RIS}-Rx link. An example of a \gls{RIS} deployment strategy is in \cite{RIS_delpoymen_Pos}, where the authors analyze a single exemplary geometry of Tx, Rx and \gls{RIS}, and devise the best relative position of the \gls{RIS} given its distance from the Tx-Rx baseline. In \cite{RIS_delpoymen_Pos2}, instead, the authors show that in case the Tx employs a very narrow beam (whose projection onto the \gls{RIS} is smaller than the \gls{RIS} size), the best \gls{RIS} deployment is close to the Rx terminal.

A leap forward in the analysis of \gls{RIS} deployment is in \cite{COV_Analysis_0,bafghi2024stochastic}, where the authors perform the coverage analysis for a \gls{RIS}-assisted wireless network, with the aid of stochastic geometry. The former two works aim to yield a stochastic expression of the coverage probability, where BSs, user equipments (UEs), blockers, and \gls{RIS}s might have different stochastic distributions. In \cite{COV_Analysis_0} it is shown that using \gls{RIS}s results in significant coverage enhancement, particularly when the density of the blockers in space is high. This outcome aligns with similar results obtained in the context of V2V communications \cite{CEMS_V2V}. In \cite{bafghi2024stochastic}, the authors also consider the effect of interference, showing that as the density of the deployed \gls{RIS}s increases, the power of the desired and interfering signals also increases. In addition to the optimal placement of the \gls{RIS}, the optimal orientation and phase configuration are analyzed in \cite{COV_OPT_2, COV_OPT_3, RIS_COV_OPT_INDOOR}. The work \cite{COV_OPT_2} extends \cite{RIS_delpoymen_Pos, RIS_delpoymen_Pos2} considering the orientation, while the authors of \cite{COV_OPT_3} deal with a coverage range maximization problem considering a \gls{STAR}. The locations and orientations of multiple indoor \gls{RIS}s to optimize coverage, specifically within shadowed regions without line-of-sight, are discussed in \cite{RIS_COV_OPT_INDOOR}.

Concerning \gls{NCR}s, the available literature is limited. For example, various deployment options for \gls{NCR}s are discussed in \cite{ncr_deployement}, showing increased coverage performance, especially for cell edge UEs. An evaluation of \gls{NCR} at the system level is reported in \cite{NCR_Interf_Makki}, which shows that the deployment of \gls{NCR} must be sensitive to interference, as far as multiple cells are concerned. 
The comparison between \gls{RIS}s and active relays, such as decode\&forward (DF) and amplify\&forward (AF), has been addressed in the first literature works, such as \cite{RIS_vs_DF} and \cite{RIS_vs_AF}. These works demonstrate that a very large \gls{RIS} is needed to match the coverage and capacity performance of DF relays, thus the interest is in comparing \gls{RIS}s and AF relays. In \cite{comparison_makki}, \gls{RIS} and \gls{NCR} are compared in a simplified illustrative scenario, showing the effect of different beamforming methods and hardware impairment models of the \gls{RIS}, as well as different parameters of the NCR, such as amplification gain.
The aforementioned works (and reference therein) provide insightful guidelines for \gls{HSRE} planning in simplified scenarios only, where the positions of BSs, UEs and \gls{RIS}s/\gls{NCR}s are selected without constraints (no buildings, no obstacles, no realistic environments). Therefore, a more realistic planning of the \gls{HSRE} must take place in real environments. In \cite{RIS_vs_NCR_Reza}, \gls{RIS}s are compared with \gls{NCR}s from both the propagation and geometric perspectives, in an urban scenario, where the constraints of buildings and installation are taken into account. The study demonstrates that the preference between \gls{RIS} and \gls{NCR} depends not only on their physical capabilities, such as the dimensions of the \gls{RIS}, the end-to-end amplification gain, and the angular separation between the \gls{NCR} panels, but also on the geometry of the environment. This work sets the ground for the effective integration of \gls{HSRE} in large-scale network planning. 

From the network planning perspective, few works are available for \gls{HSRE}. \textcolor{black}{However, comprehensive considerations for network planning are crucial. A recent study \cite{emil_planning} has shown that simply increasing the number of RIS devices does not always lead to proportional performance gains in network coverage, particularly when deployment costs are considered. In \cite{makki_planning}, the performance of a dense \gls{IAB} network in the presence of \gls{SRE} components is evaluated, demonstrating the efficiency of these components in some specific scenarios, such as outdoor sport events. This works calls for an efficient planning approach, but does not focus on it.}

In \cite{Massa_Planning}, a system-by-design framework is presented to determine the optimal locations and layouts of the minimum number of synthesized passive static metasurfaces to meet coverage requirements in a predefined region of interest. The framework considers a set of candidate sites in realistic urban propagation scenarios for metasurface installation and employs a binary genetic algorithm to decide on the installation sites. However, \gls{RIS}s or \gls{NCR}s are not considered. Other planning strategies that consider both \gls{RIS}s and relays are in \cite{RIS_RS_Eugenio}, again in a simplified scenario without installation constraints or realistic environment. In \cite{Eugenio2}, a novel mathematical formulation of coverage planning is presented in the \gls{RIS}-assisted communication problem, which leads to increased throughput and coverage. In \cite{RIS_RS_Eugenio}, an optimal IAB-aided RAN in m\gls{mmW} is presented, with the aim of enhancing the reliability of the network by determining the best positions and configurations for \gls{NCR} and \gls{RIS}. The approach focuses on maximizing the angular separation between the different paths reaching a given candidate site, as well as minimizing the average link length to mitigate blockage \cite{DynamicBlockage}. The numerical results confirm that joint installation of both \gls{NCR} and \gls{RIS} significantly increases reliability compared to the use of either one alone.

\begin{figure*}[t!]
    \centering
    \subfloat[]{\includegraphics[width=0.33\textwidth]{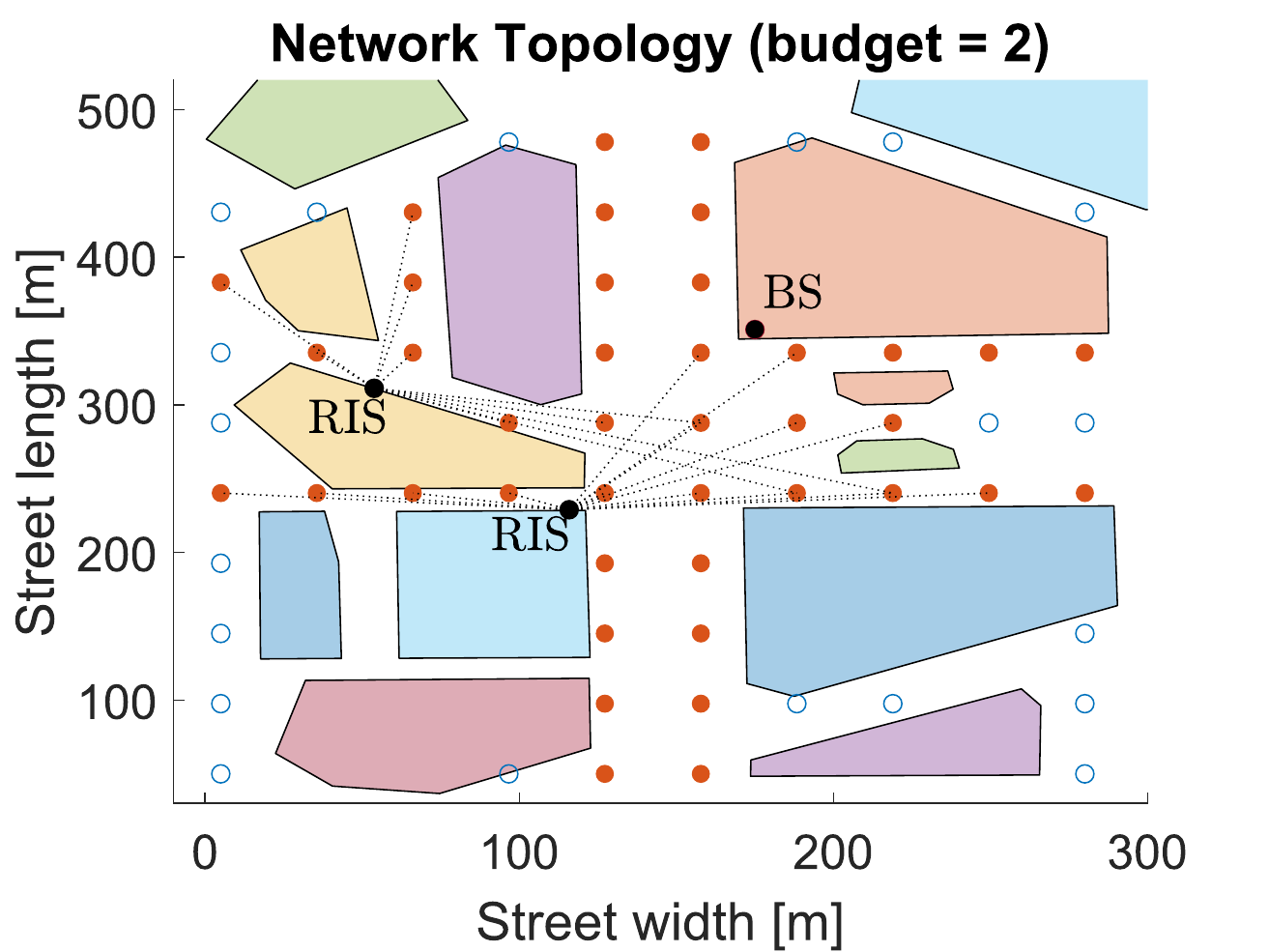}} 
    \subfloat[]{\includegraphics[width=0.33\textwidth]{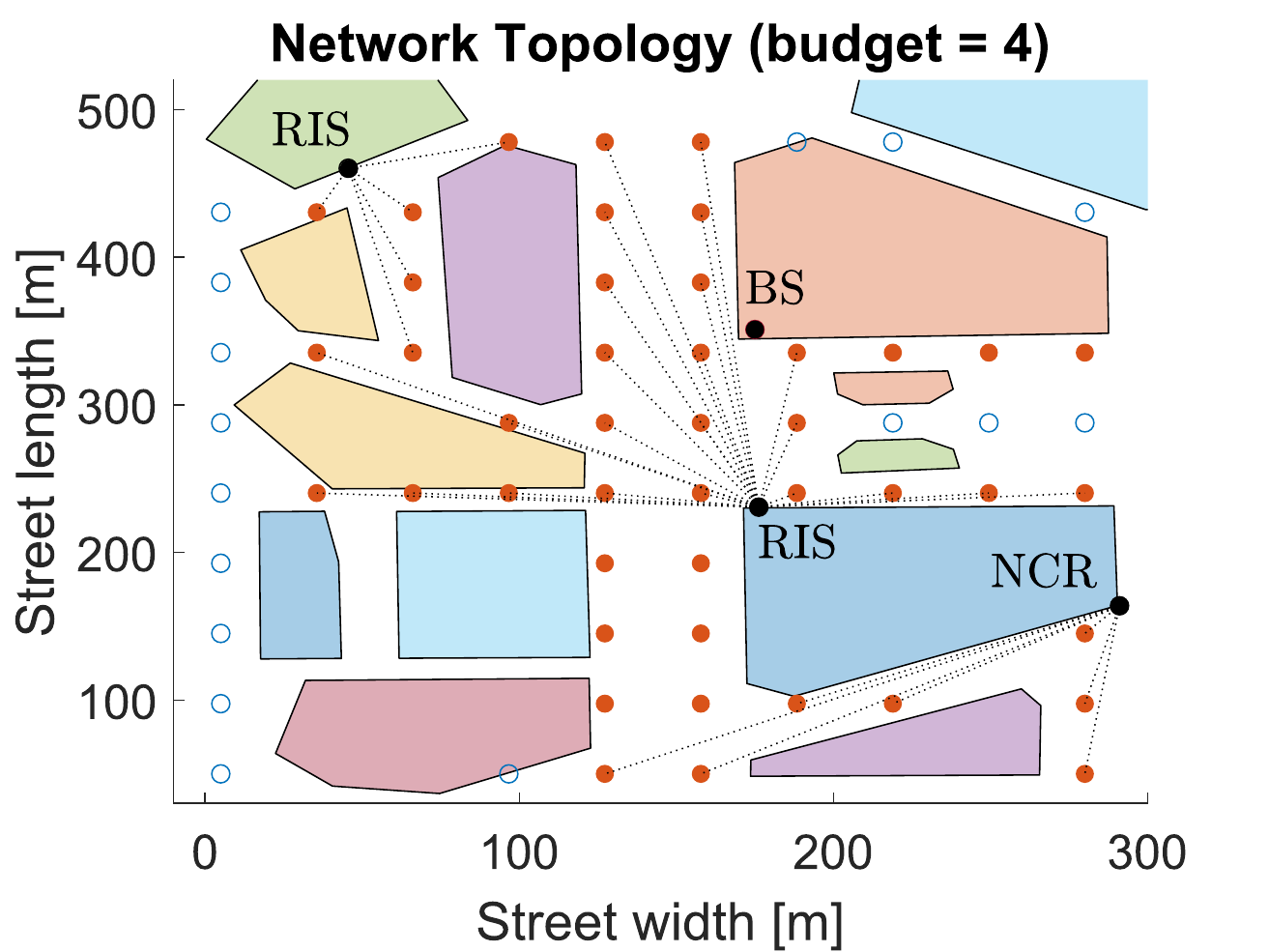}}
    \subfloat[]{\includegraphics[width=0.33\textwidth]{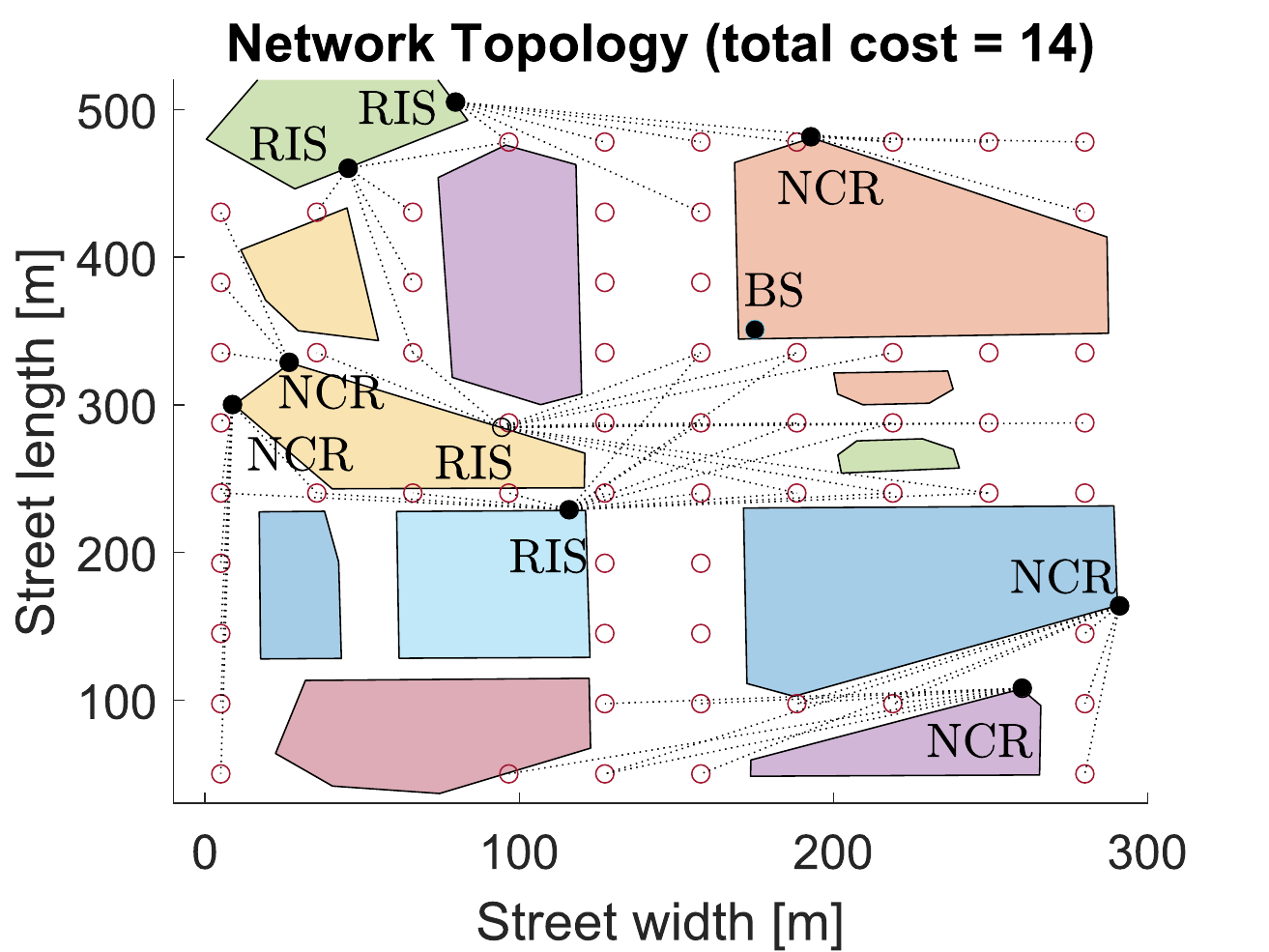}}
    \caption{Network topology with a single fixed BS, \gls{RIS} and/or \gls{NCR} devices, achieved with (a) MBCC model, given fixed budget of  $B=2$ units, (b) MBCC model, given fixed budget of $B=4$ units, and (c) FCMC model. The price of each \gls{RIS} is assumed to be 1 unit and each \gls{NCR} is 2 units.  }\label{fig:motivation}
\end{figure*}
\subsection{Contribution}

To address current gaps in network planning for 6G, our previous work \cite{planning_conf_arxiv} introduced the \gls{FCMC} model, which aims to minimize deployment costs while ensuring full area coverage using \gls{RIS} and \gls{NCR} components. Building on this foundation, in this paper, we provide a comprehensive approach to network planning within the \gls{HSRE}, expanding the \gls{FCMC} model by incorporating additional device types, such as \gls{STAR} and \gls{3SNCR}, along with a new optimization approach, the \gls{MBCC} model, which balances coverage with budget limitations. Using realistic urban maps, precise blockage model and installation constraints, this expanded framework allows for a more accurate analysis of network planning strategies that address both full-coverage and budget-constrained scenarios, making it applicable to diverse urban deployment conditions.

The list of detailed contributions is as follows:
\begin{itemize}
    \item We consider the precise channel models of \gls{RIS}s and \gls{NCR}s, accounting for all propagation characteristics, and their physical/geometric limitations. We used a realistic map of the city of Milan for planning purposes. Each \gls{HSRE} component, given its characteristics, conforms to the layout of the environment. Furthermore, we use cost models in our numerical results for \gls{HSRE} components that scale with the settings of the components.
    \item In addition to already prototyped/deployed components such as \gls{RIS}s and \gls{NCR}s, we also consider the simultaneous usage of next-generation \gls{HSRE} components such as \gls{STAR} and \gls{3SNCR} and evaluate their feasibility and effects on performance metrics. 
    \item We provide two mathematical optimization models to minimize the average total cost, given the full coverage of the area considered, namely the \gls{FCMC} model, and to maximize the coverage percentage, given a fixed budget, namely the \gls{MBCC} model. The extensive numerical analysis performed by applying the proposed models, allowed us to derive important guidelines on how to configure
and deploy \gls{HSRE} components.
\end{itemize}

\subsection{Organization}
This paper is organized as follows. Section \ref{sect:motivation} motivates deploying \gls{HSRE} in network planning. Section \ref{sect:SRE} introduces the system model. In Section \ref{sect:technology} a detailed description of the components of the \gls{SRE} with an estimated cost model is discussed. In section \ref{sect:optimization}, we formalize the coverage optimization problem, proposing two distinct optimization models. Section \ref{sect:results} presents the numerical results and provides insights into deployment strategies, device configurations, and cost analysis. Finally, Section \ref{sect:conclusion} concludes the paper.

 \subsection{Notation}\label{sec:notation}

The bold upper- and lowercase letters represent the matrices and the column vectors, respectively. The $(i,j)$-th entry of the matrix $\mathbf{A}$ is denoted by $[\mathbf{A}]_{ij}$. Transpose and conjugate transpose of matrix $\mathbf{A}$ are represented by $\mathbf{A}^T$ and $\mathbf{A}^H$, respectively. The identity matrix of size $n$ is written as $\mathbf{I}_n$. A complex-valued scalar random variable is denoted by $z \sim \mathcal{CN}(0, \sigma^2)$, where $\mathcal{CN}$ represents the circularly symmetric complex Gaussian distribution. Value of a variable $x$ in dB scale is denoted as $[x]_{\textrm{dB}}$. The operator $\mathbb{E}[\cdot]$ denotes the expectation, while $\mathbb{C}$ and $\mathbb{R}$ represent the sets of complex and real numbers, respectively. The symbol $\|\cdot\|_F$ represents the Frobenius norm.

\section{Motivation\label{sect:motivation}}

Network planning in the \gls{SRE} is affected by many parameters. These parameters include configuration and field of view of the devices, static blockers (buildings in our case), geometry of the streets, accurate propagation details of the devices, and overall, the environment. More importantly, the planning depends on how much each of the devices would cost (deployment, scaling, maintenance, or power consumption), and how much budget is available, and what is the optimization goals with related constraints. In this example, Fig. 1 shows a sample network topology for a section of Milan. In this scenario, the configuration and position of a single \gls{BS} are predefined, while the position and orientation of the \gls{SRE} devices, \gls{RIS}s and the \gls{NCR}s, are the focus of optimization. \gls{NCR}s have an end-to-end amplification gain of 95 dB, and \gls{RIS}s employ $100 \times 100$ half-wavelength meta-atoms. The configurations and details of the devices and the optimization model will be discussed exhaustively later in the following sections. Here, each \gls{RIS} costs 1 units and each \gls{NCR} costs 2 units. Fig. \ref{fig:motivation} (a) and Fig. \ref{fig:motivation} (b), are examples, showing the network topology for the case where we use the \gls{MBCC} optimization model, whose goal it is to maximize the coverage, when the total available budget is constrained to 2 and 4 units, respectively. Here, the empty circles are the \gls{tp} that are not covered, while the filled red circles are the \gls{tp}s that are covered, either via a relay (\gls{RIS} or NCR), or directly by the BS, or both. The dashed lines show the coverage of \gls{tp} by the corresponding relaying device. It can be seen that, by changing the available budget, not only the devices used and their number will change, but also their position and orientation will be affected. Increasing the available budget, more \gls{tp}s can be covered. Instead, in Fig. \ref{fig:motivation} (c) we use the \gls{FCMC} optimization model, whose goal is to minimize the total cost, while guaranteeing full coverage of all \gls{tp}s and, thus, the budget is not limited. It can be seen that the full coverage requirement will not only drastically increase the total cost but will also change the entire network topology, showing how different optimization goals affect planning in \gls{SRE}.

This example shows how various parameters will be decisive for network planning optimization. Therefore, the goal of the paper is to make a comprehensive consideration of all the effective parameters, as precise as possible, and to propose appropriate optimization models that suit different requirements and goals. 

\section{Smart Radio Environment Model\label{sect:SRE}}

Consider the downlink communication shown in Fig. \ref{fig:motivation}, which shows two possible connections, direct and relayed, between a BS and all \gls{tp} (that stands for possible UEs). The locations of BS, relay, and UE in space, described in a global coordinate system, are indicated as $\mathbf{p}_{BS} = [x_{BS}, y_{BS}, z_{BS}]^\mathrm{T}$, $\mathbf{p}_R = [x_R, y_R, z_R]^\mathrm{T}$, $\mathbf{p}_{UE} = [x_{UE}, y_{UE}, z_{UE}]^\mathrm{T}$, respectively. The set of all potential UE locations is given by $\mathcal{P}=\{\mathbf{p}_{UE}\}$. The BS and the UEs have an antenna array of $N_t$ and $N_r$ elements, respectively.

\begin{figure*}[t!]
    \centering
    \subfloat[][\small RIS]{\includegraphics[height=0.2\textwidth]{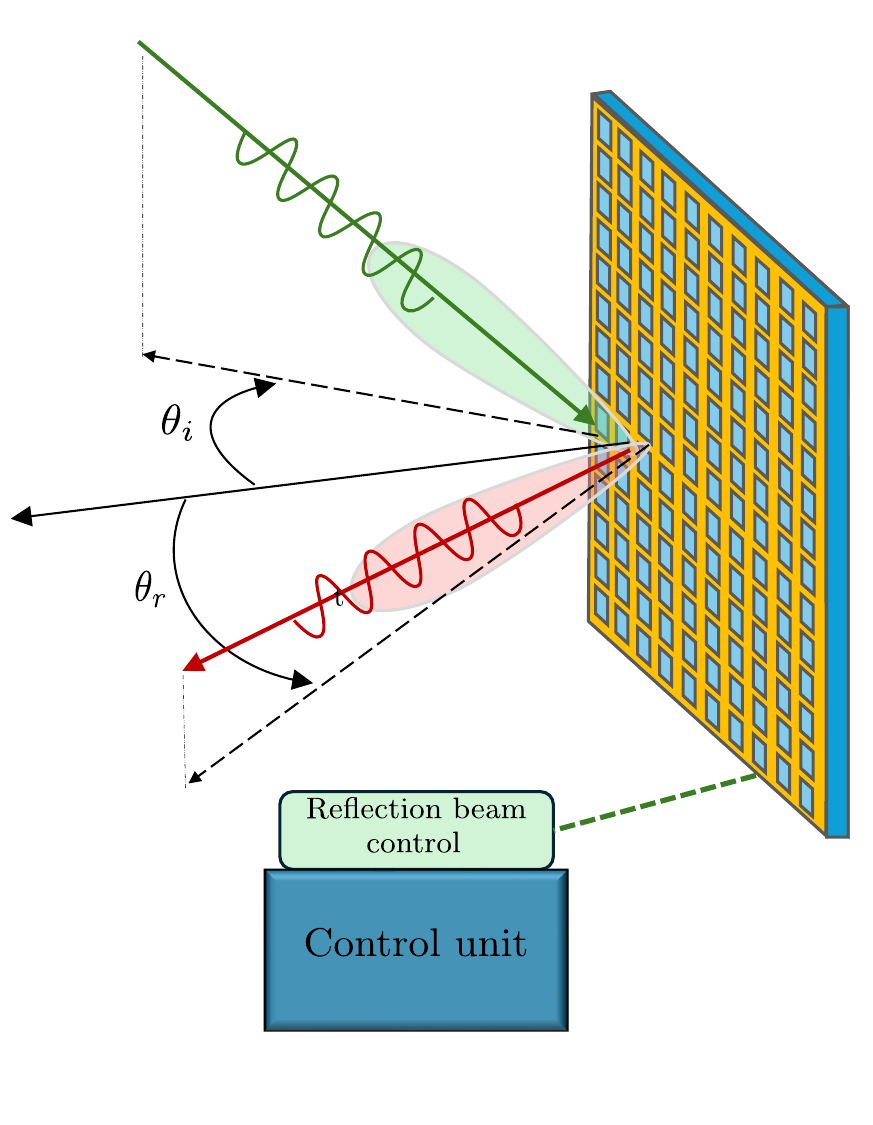}\label{fig:RIS}}\quad
    \subfloat[][\small \gls{STAR}]{\includegraphics[height=0.2\textwidth]{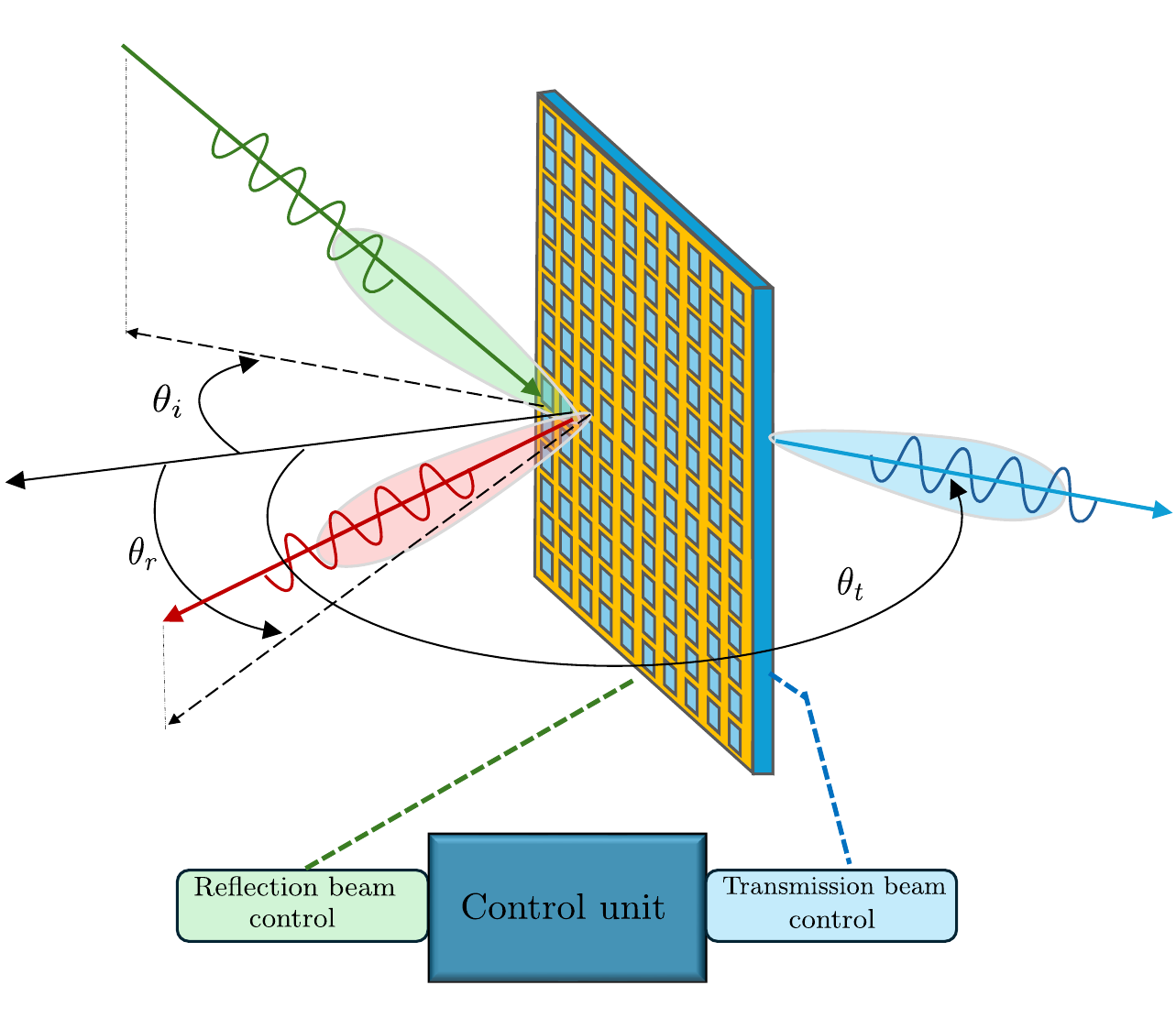}\label{fig:STAR}}\quad
    \subfloat[][\small NCR]{\includegraphics[height=0.2\textwidth]{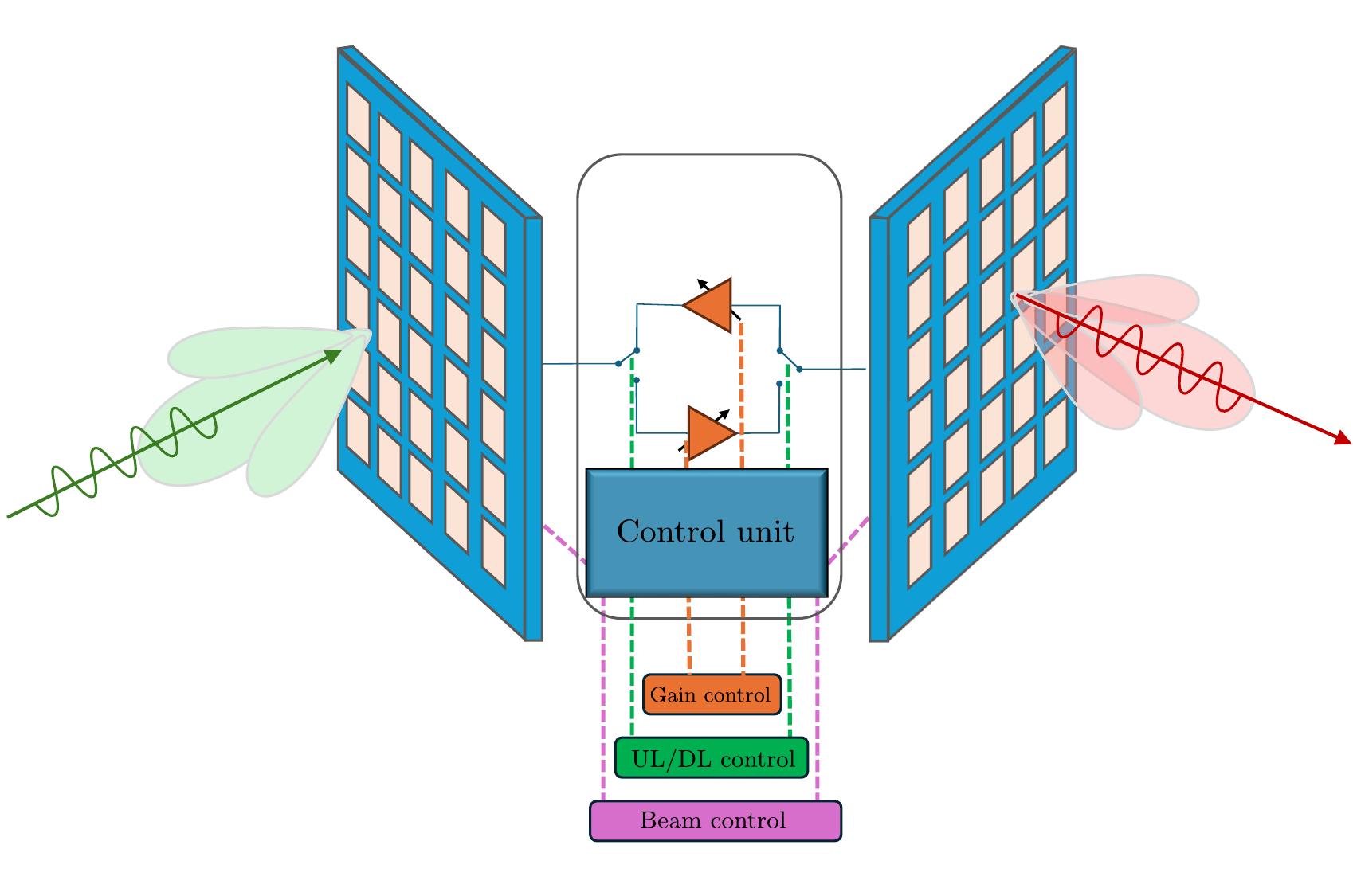}\label{fig:NCR}}\quad\subfloat[][\small \gls{NCR} vs \gls{3SNCR}]
    {\includegraphics[height=0.2\textwidth]{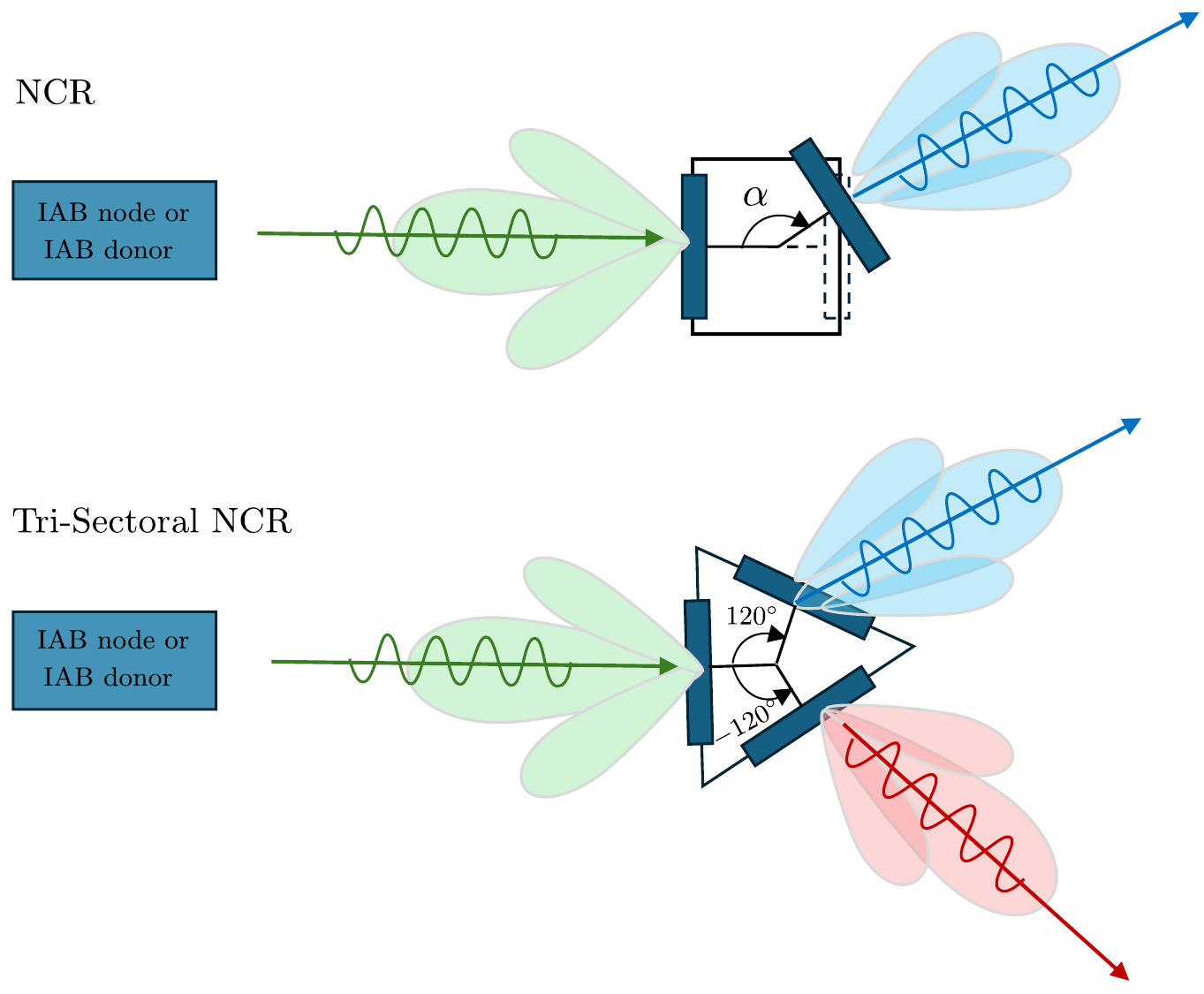}\label{fig:3NCR}}
    \caption{Illustrative scheme of different \gls{SRE} components}
\end{figure*}

The relay unit can be either an intelligent metasurface or an NCR. Two types of intelligent metasurfaces are considered: \gls{RIS} and \gls{STAR}, depicted in Figs. \ref{fig:RIS} and \ref{fig:STAR}, both using $M$ meta-atoms. For \gls{NCR}s, two configurations are examined: two-panel and \gls{3SNCR}, shown in Fig. \ref{fig:NCR} and \ref{fig:3NCR}, respectively. 
In the two-panel \gls{NCR} configuration, the first panel faces the BS, while the second panel faces the coverage area, oriented at an angle $\alpha$ relative to the first panel. Each panel is equipped with $N_p$ antenna elements. In the three-panel \gls{NCR} configuration, the first panel, which contains $N_p$ antenna elements, is oriented toward the BS, while the other two panels, each with $N_p/2$ antenna elements to comparable complexity. They are positioned so that the angle between each pair of panels is $120^\circ$.
%
The antenna arrays have half-wavelength spacing at the carrier frequency $f_0=c/\lambda$ (for the speed of light $c$ and wavelength $\lambda$), while the intelligent metasurface meta-atoms are spaced at $\lambda/4$ intervals.

\subsection{Signal Model}\label{subsect:signal_model}

Let $s \in \mathbb{C}$ be the complex symbol to be transmitted, such that $\mathbb{E}[s^*s] = \sigma_s^2$, where $\sigma_s^2$ denotes the transmitted power. The transmitted signal is expressed as
\begin{equation}\label{eq:txSignal}
    \mathbf{x} = \mathbf{f}\,s,
\end{equation}
where $\mathbf{f}$ denotes the precoding vector, such that $\|\mathbf{f}\|_\mathrm{F}^2 = N_t$. The signal transmitted in \eqref{eq:txSignal} can be received by the UE directly $\mathbf{r}^\mathrm{d}$, through \gls{NCR} $\mathbf{r}^\mathrm{NCR}$, or via an intelligent metasurface $\mathbf{r}^\mathrm{IM}$. This can be expressed as
\begin{equation}\label{eq:rxTotal}
    y = \mathbf{w}^\mathrm{H} \left(\mathbf{r}^\mathrm{d} + \sum_{i\in \mathcal{I}} \mathbf{r}_i^\mathrm{IM} + \sum_{j\in \mathcal{J}} \mathbf{r}_j^\mathrm{NCR}\right) + \mathbf{w}^\mathrm{H} \, \mathbf{n},
\end{equation}
where $\mathbf{n}\sim \mathcal{CN}\left(\mathbf{0}, \sigma_n^2 \, \mathbf{I}_{N_r}\right)$ denotes the additive white Gaussian noise, $\mathbf{w}\in \mathbb{C}^{N_r \times 1}$ is the combining vector, and $\mathcal{I}, \mathcal{J}$ denote the set of deployed intelligent metasurfaces and \gls{NCR}s, respectively.

\subsubsection{Direct Signal Model}
The received signal model for the direct link can be expressed as
\begin{equation}\label{eq:rxDirect}
    \mathbf{r}^\mathrm{d}  =  \mathbf{H}^\mathrm{d} \, x
\end{equation}
with $\mathbf{H}^\mathrm{d} \in \mathbb{C}^{N_r \times N_t}$ being the MIMO direct channel.

\subsubsection{Relayed Signal Model via Intelligent Metasurface}
The received signal model for the relayed link through an intelligent metasurface can be expressed as
\begin{equation}\label{eq:rxIM}
    \mathbf{r}^\mathrm{IM} = \mathbf{H}_o^\mathrm{IM} \, \left(\sqrt{\beta_r} \, \boldsymbol{\Phi}_r + \sqrt{\beta_t} \, \boldsymbol{\Phi}_t\right) \, \mathbf{H}_i^\mathrm{IM} \, \mathbf{x},
\end{equation}
where $\beta_r$ and $\beta_t$ denote the reflection and transmission coefficient of the intelligent metasurface, respectively, such that $\beta_r + \beta_t \leq 1$, with $\beta_t = 0$ in case the intelligent metasurface is an \gls{RIS}, and $\beta_t > 0$ for \gls{STAR}. The reflection and transmission coefficient matrices are denoted by $\boldsymbol{\Phi}_r \in \mathbb{C}^{M\times M}$ and $\boldsymbol{\Phi}_t \in \mathbb{C}^{M\times M}$, $\mathbf{H}_i^\mathrm{IM} \in \mathbb{C}^{M\times N_t}$ and $\mathbf{H}_o^\mathrm{IM} \in \mathbb{C}^{N_r\times M}$ are the input/output channel matrices between the BS, the intelligent metasurface, and the UE. The reflection/transmission matrix $\boldsymbol{\Phi}$ in \eqref{eq:rxIM} is diagonal with entries defined as
\begin{equation}\label{eq:reflectingmatrix}
    \boldsymbol{\Phi} = \text{diag}\left( e^{j\phi_{1}},...,e^{j\phi_{m}},..., e^{j\phi_{M}}\right),
\end{equation}
where $\phi_{m}$ denotes the phase applied at the $m$-th element. The phase shifts are assumed to be designed as optimal for every Rx position \cite{CIRS}, given the nature of network planning.  

\subsubsection{Relayed Signal Model via Network Controlled Repeater}
Since the \gls{NCR} is an active device, the model of the received signal includes the amplified noise. The signal received by the \gls{NCR} can be expressed as
\begin{equation}\label{eq:rxNCR}
    \mathbf{z}^\mathrm{NCR} = \mathbf{H}_i^\mathrm{NCR} \, \mathbf{x} + \mathbf{v}
\end{equation}
where $\mathbf{v} \in \mathbb{C}^{N_p \times 1} \sim \mathcal{CN}\left(\mathbf{0}, \sigma_v^2 \, \mathbf{I}_{N_p}\right)$ denotes the noise at the NCR, and $\mathbf{H}_i^\mathrm{NCR} \in \mathbb{C}^{N_p \times N_t}$ is the channel matrix between the BS and the NCR. The signal $\mathbf{z}^\mathrm{NCR}$ in \eqref{eq:rxNCR} is amplified and forwarded towards the UE. The signal received by the UE through the \gls{NCR} can be expressed as
\begin{equation}
    \mathbf{r}^\mathrm{NCR} = \sqrt{g} \, \mathbf{H}_o^\mathrm{NCR} \, \mathbf{Q} \, \mathbf{z}^\mathrm{NCR} 
\end{equation}
where $g$ denotes the amplification gain, $\mathbf{H}_o^\mathrm{NCR} \in \mathbb{C}^{N_r \times N_p}$ denotes the channel matrix between the \gls{NCR} and the UE. The relaying matrix $\mathbf{Q} \in \mathbb{C}^{N_p \times N_p}$ operates a phase shift between the received signal and the forwarded one. In case the \gls{NCR} has two panels, the relaying matrix  can be expressed as
\begin{equation}
    \mathbf{Q} = \mathbf{u} \, \mathbf{b}^\mathrm{H}
\end{equation}
where $\mathbf{u} \in \mathbb{C}^{N_p \times 1}$ and $\mathbf{b} \in \mathbb{C}^{N_p \times 1}$ denote the forwarding and receiving beamformers, respectively. In the case, the \gls{NCR} has three panels, the relaying matrix is expressed as
\begin{equation}
    \mathbf{Q} = \mathbf{U} \, \boldsymbol{\xi} \, \mathbf{b}^\mathrm{H}
\end{equation}
where $\boldsymbol{\xi} \in \mathbb{B}^{2\times 1}$ is a selection vector that allocates the incoming stream to one of the output panels, and $\mathbf{U} \in \mathbb{C}^{N_p \times 2}$ is a block diagonal matrix, defined as
\begin{equation}
    \mathbf{U} = \begin{bmatrix}
        \mathbf{u}^{(1)}  & \mathbf{0}\\
        \mathbf{0}      & \mathbf{u}^{(2)}
    \end{bmatrix}
\end{equation}
with $\mathbf{u}^{(1)} \in \mathbb{C}^{N_p/2 \times 1}$ and $\mathbf{u}^{(2)} \in \mathbb{C}^{N_p/2 \times 1}$ denoting the forward beamforming vectors to each of the output panels, and $\mathbf{0}$ is an $N_p/2 \times 1$ zeros vector.

\subsection{Channel model}\label{subsect:channel_model in presence of RIS}

This paper assumes a block-fading channel model with independent fading among direct $\mathbf{H}^\mathrm{d}$, forward $\mathbf{H}_i$, and backward $\mathbf{H}_o$ channels from \gls{NCR} and intelligent metasurfaces. Considering the challenging propagation conditions at mmWave frequencies, we have adopted the Saleh-Valenzuela cluster-based model \cite{Molisch}, in this study. The impulse response of any channel in the model being considered can be expressed as
\begin{equation}\label{eq:channelModel}
    \mathbf{H} = \sum_{p=1}^{P} \frac{\alpha_p}{\sqrt{P}}\, \varrho_r(\vartheta_p^r, \varphi_p^r) \varrho_t(\vartheta_p^t, \varphi_p^t) \mathbf{a}_r(\vartheta_p^r, \varphi_p^r)\mathbf{a}_t(\vartheta_p^t, \varphi_p^t)^\mathrm{H}, 
\end{equation}
where $P$ is the number of paths, $\alpha_p$ denotes the scattering amplitude of the $p$-th path, and $\mathbf{a}_t(\vartheta_p^t, \varphi_p^t)\in\mathbb{C}^{N_t\times 1}$ and $\mathbf{a}_r(\vartheta_p^r, \varphi_p^r)\in\mathbb{C}^{N_r\times 1}$ denote the Tx and Rx array response vectors for the $p$-th path, functions of the Tx and Rx pointing angles $(\vartheta_p^t, \varphi_p^t)$ and $(\vartheta_p^r, \varphi_p^r)$. For convenience, the channel model incorporates the element radiation patterns $\varrho(\vartheta_p, \varphi_p)$, which is defined according to \cite{3GPP} for BS, UE, and NCR, while for the intelligent metasurface, it is defined as in \cite{CIRS}.

The array factor $\mathbf{a}(\vartheta, \varphi)$ in \eqref{eq:channelModel}, assuming a uniform planar array, can be expressed as
\begin{equation}
    \mathbf{a}(\vartheta, \varphi) = [e^{j\mathbf{k}(\vartheta, \varphi)^\mathrm{T} \boldsymbol{\nu}_1}, \cdots, e^{j\mathbf{k}(\vartheta, \varphi)^\mathrm{T} \boldsymbol{\nu}_L}]
\end{equation}
where $\mathbf{k}(\vartheta, \varphi) \in \mathbb{R}^{3 \times 1}$ is the wave vector, defined as
\begin{equation}
    \mathbf{k}(\vartheta, \varphi) = \frac{2\pi}{\lambda}[\cos(\varphi)\cos(\vartheta), \cos(\varphi)\sin(\vartheta), \sin(\varphi)]^T,
\end{equation}
and $\boldsymbol{\nu}_\ell = [x_\ell, y_\ell, z_\ell]$ is the position of the $\ell$th element of the considered ULA array, expressed in local coordinates. 

Having calculated direct or relay channels, instantaneous \gls{SNR} $\gamma_0$ through every link, can be calculated similarly to \cite{RIS_vs_NCR_Reza}. The static blockage here is deterministic as we use the actual map of the buildings to form the scenario. The dynamic blockage model is taken from \cite{DynamicBlockage}, used to calculate the long-term \gls{SNR}. For instance, for the direct path, the long-term \gls{SNR} is calculated as \cite{RIS_vs_NCR_Reza}
\begin{align}
    &\overline{\gamma}^\textrm{BS} = \mathrm{P}^{\textrm{BS}}_B\,\gamma^{\textrm{BS}} + (1-\mathrm{P}^{\textrm{BS}}_B)\gamma_{0}^\textrm{BS},\label{eq:LT_SNR_direct}
\end{align}
where $\mathrm{P}^{\textrm{BS}}_B$ is the blockage probability of the direct paths, $\gamma_{0}^\textrm{BS}$ is the \gls{SNR} when blockage does not happen, while ${\gamma}^\textrm{BS}$ is the \gls{SNR} considering the penetration loss or knife edge diffraction \cite{RIS_vs_NCR_Reza} \cite{3GPPTR38901}. The long-term \gls{SNR} of the relayed links can be calculated similarly. These long-term \gls{SNR}s will be used by the optimization models in the next section. The dynamic blockage probability $\mathrm{P}_B$ depends on the length of the considered link, as well as other parameters such as blockers' density, velocity, and dimensions \cite{DynamicBlockage}.

\section{\gls{SRE} devices' technologies AND Costs}\label{sect:technology}
\gls{RIS} (Fig.~\ref{fig:RIS}) are typically constructed using metasurfaces, which are thin, planar structures composed of sub-wavelength-sized elements. The primary components of an \gls{RIS} include the sub-wavelength elements and their feed system, each controlled by PIN diodes or varactors, the control circuitry and microprocessors, necessary for dynamic reconfiguration \cite{SRE}. Based on these, the total cost of a RIS can be divided to the constant deployment and components' cost and variable components' cost scaling linearly with the number of \gls{RIS} meta-atoms.

 One main drawback of \gls{RIS} is that it can only cover a maximum of half of the plane, that is, the source and destination must be in front of the \gls{RIS} to be served. A novel alternative to the reflective RIS, is the \gls{STAR}. The dual functionality of \gls{STAR} is achieved instead through specialized hardware, often constructed using graphene or multilayered metal patches separated by a polyimide substrate \cite{STAR_Bifunctional} or dielectrics \cite{Star_FullDimension}. In the initial prototypes such as in \cite{RFocus} or \cite{IOS_BeyondRIS} the independent control of reflection and refraction was not possible and the power is split on both sides of the surface. In more advanced hardware models, more degrees of freedom are available to guarantee the manipulation of the amplitude and phase shift, simultaneously, for transmitted and reflected waves \cite{STAR_FullSpace}. Independent control for transmission and reflection is enabled by using separate controlling components for the elements in both sides, to manipulate the electric and magnetic currents. This would translate to at least twice more electronic components \cite{StarGaldi}.

\gls{NCR}s (Fig.~\ref{fig:NCR}) are controlled by base stations to enhance network coverage and performance. Unlike traditional RF repeaters, \gls{NCR}s act as extensions of the base station, using beamforming and timing synchronization to effectively manage interference and energy consumption. They are transparent to mobile devices and adhere to \gls{3gpp} standardized procedures for authentication and authorization \cite{pivotal5g}. \gls{NCR} employs two panels, each with a \gls{UPA}. Both panels can perform beamforming, while the \gls{NCR} decides also the amplification gain, and time division duplexing, given the network layer information provided by the BS, an \gls{IAB} donor or an \gls{IAB} node, if any \gls{IAB} architecture is deployed (Fig. \ref{fig:NCR}). The angular separation between the two panels, to avoid loop-back interference, is one of the main downsides of the NCR, as one panel must face the BS, while the other panel has a limited \gls{FoV}. The key cost components of \gls{NCR}s include beamforming antennas, RF amplifiers, TDD switching units, and the control unit that manages the beamforming and amplification control. Since for a \gls{3SNCR}, we consider that the second and third panels have half the number of antennas of a regular NCR, it can be considered that their cost is similar. However, it is expected that an \gls{NCR} costs much more than a \gls{RIS}, given that it has power amplifiers and continuous power consumption. Thus, it can be argued that the total cost of a \gls{NCR} would include site installation and initial component costs (i.e., part of the CAPEX), as well as the cost of operating and maintaining the device (i.e., part of the OPEX). 

Based on the above-mentioned considerations, we presume a cost model for the devices as listed in Table~\ref{tab:SimParam_2}. Similarly to \cite{Eugenio2}, these cost models will be used for networks planning optimization.

Remark: The cost models applied here are rational estimates based on component needs and operational demands. For clarity, prices are presented in units, with a $100\times 100$ RIS taken as the reference, costing 1 unit. Other device prices are scaled relatively to this baseline, emphasizing that in our optimization, it is the relative, not absolute, prices that influence the outcomes. Although actual costs will depend on market conditions and production scales, our framework can seamlessly accommodate updates to these models as vendor data becomes available. Importantly, while variations in the cost model may affect quantitative results, they do not alter the optimization approach, which is agnostic to specific pricing assumptions.

\begin{table}[t!]
    \centering
    \footnotesize
    \caption{Device costs scaling with configurations .}
    \begin{tabular}{l|c|c}
    \toprule
        \textbf{Parameter} &  \textbf{Symbol} & \textbf{Value(s)}\\
        \hline
         RIS size & $\ M $  & $ 50\times50 - 300\times 300$ \\
         NCR gain in dB & $[g]_{dB} $  & $ 30-70$ \\
         RIS cost per unit cell& $\ O ^{ris}_c $  & $ 6 \times 10^{-5}$ \\
                NCR deployment cost & $O ^{ncr}_d$  & $0.8$\\
                 NCR cost per dB of Gain & $O ^{ncr}_g$  & $4\times 10^{-2}$\\
                 RIS price & $\ O ^{ris} $  & $ O ^{ris}_{d} + \ O ^{ris}_{c} \times M$ \\
        
        STAR RIS price & $O ^{star}$  & $2\times \ O ^{ris}$\\
        NCR price & $O ^{ncr}$  & $O ^{ncr}_{d} +  O ^{ncr}_{g} \times [g]_{\textrm{dB}}$\\
         3SNCR price & $O ^{3sncr}$  & $O ^{ncr}$\\
       
        \bottomrule
    \end{tabular}
    \label{tab:SimParam_2}
\end{table}
%

%

\section{Coverage Optimization\label{sect:optimization}}

mmWave \gls{SRE}-based networks are heavily based on deployment geometry, directly impacting their performance, as indicated in previous planning studies~\cite{RIS_RS_Eugenio}. Given large-scale deployments due to the low cost of these devices, understanding the overall impact at the system level is crucial. To accurately assess network performance when these devices are integrated, we employ a \gls{milp} optimization approach. This method guarantees an optimal deployment layout, ensuring that the performance insights derived represent the best possible outcomes for networks that use these devices. In particular, we employ two different optimization models reflecting complementary network planning objectives, where either the cost is minimized or the coverage is maximized. 

\subsection{System Model}
\newcommand{\R}{\mathcal{R}}
\newcommand{\N}{\mathcal{N}}

Consider a geographic area where the coverage of a pre-deployed BS has to be enhanced by the installation of \gls{SRE} devices. In particular, let $\mathcal{C}$ be the set of \gls{SRE} \glspl{cs}. Any $c \in \mathcal{C}$ represents the position in the geographic area where an \gls{SRE} device can be installed.

As discussed in Sect.~\ref{subsect:signal_model}, this work considers 4 different types of \gls{SRE} devices that present a diverse set of performance metrics and costs. As such, the \gls{milp} planning approach should be able to select the best \gls{SRE} device that maximized the planning objective in every active \gls{cs}. This is mathematically modeled by employing a set $\mathcal{D}$, which represents the collection of all available \gls{SRE} technologies to install. The set $\mathcal{D}$ not only allows optimizing network planning by selecting different technologies (\gls{RIS} or NCR), but also allows the configuration of the selected technology to be further optimized, i.e., \gls{RIS} size or end-to-end \gls{NCR} gain. Therefore, its cardinality may be larger than the number of \gls{SRE} device types.

Let $\mathcal{T}$ be the set of \glspl{tp}, whose location corresponds to any spatial sampling of the geographic area and can be considered as the positions where the coverage has to be evaluated. We consider a particular \gls{tp} $t\in \mathcal{T}$ covered when the \gls{SNR} measured at that location is greater than a parametric threshold $\Gamma$. Representing the minimum guaranteed \gls{SNR} that the optimization will enforce in every \gls{tp}, $\Gamma$ is chosen before the planning optimization phase and influences the final network topology.

Nonetheless, the measured \gls{SNR} is strongly related to the network topology and depends on the BS and the active \gls{SRE} devices. In particular, any \gls{tp} $t \in T$ can be served directly by the BS or indirectly through any \gls{SRE} device. Let $\overline{\gamma}^\text{BS}_t$ be the long-term \gls{SNR} measured at \gls{tp} $t \in \mathcal{T}$ limited only to the BS contribution. $\overline{\gamma}^\text{BS}_t$ is computed according to the channel model derivations in Sect.~\ref{subsect:channel_model in presence of RIS}. Expressing the minimum \gls{SNR} requirements into a logical operator, we define the boolean direct link activation parameter 
\begin{equation}
  \Delta_t^\text{BS}=\left\{
  \begin{array}{@{}ll@{}}
    1, & \text{if}\ \overline{\gamma}_t^\text{BS} \geq \Gamma \\
    0, & \text{otherwise.}
  \end{array}\right.
  \label{eq:bs_act_param}
\end{equation}
In other words, $\Delta_{t}^\text{BS}$ is an optimization parameter with value 1 if the BS can guarantee an \gls{SNR} over threshold for the \gls{tp} $t \in \mathcal{T}$, and 0 otherwise.

In a similar fashion, we can employ the channel model derivations in Sec.~\ref{subsect:channel_model in presence of RIS} to measure the \gls{SNR} in every \gls{tp} for any \gls{SRE} technology active in any of the \glspl{cs}, and then compare it with the minimum \gls{SNR} $\Gamma$. We call this quantity $\overline{\gamma}_{t,c}^{d}$, representing the \gls{SNR} measured at \gls{tp} $t\in \mathcal{T}$ limited to the contribution of an \gls{SRE} device of technology $d \in \mathcal{D}$ installed in \gls{cs} $c \in \mathcal{C}$.

The result of the comparison operation can be logically expressed by the boolean \gls{SRE} link activation parameter 
\begin{equation}
  \Delta_{t,c}^d=\left\{
  \begin{array}{@{}ll@{}}
    1, & \text{if}\ \overline{\gamma}_{t,c}^{d} \geq \Gamma \\
    0, & \text{otherwise}
  \end{array}\right.
  \label{eq:ris_act_param}
\end{equation}

Notice that $\Delta_t^\text{BS}$ and $\Delta_{t,c}^d$ are called link activation parameters because they inform the optimization model on the possibility of activating a coverage link between the BS or any installed \gls{SRE} device and a particular \gls{tp}. Additionally, these parameters can be used to further restrict the coverage link activation. For instance, some $\Delta_t^\text{BS}$ and $\Delta_{t,c}^d$ can be fixed to 0 in case of line of sight obstructions, or to consider other geometric or geographic restrictions.

The proposed model takes into account the cost of employing \gls{SRE} devices during the network planning optimization phase. In particular, we define as $O^{d}$ the cost of installing an \gls{SRE} device of technology $d \in \mathcal{D}$ in the geographic area. As it will be shown in the remainder of this section, the planned network topology can be restricted not to exceed an overall budget, modeled through the parameter $B$.

\subsection{Full Coverage with Minimum Cost (FCMC)}
Building on the system model described above, we now define a \gls{milp} formulation for the problem of planning an SRE-empowered wireless network. In particular, the goal of the planning problem is to minimize the cost of the network while guaranteeing minimum coverage (that is, \gls{SNR} above the threshold $\Gamma$) in all \gls{tp} of the geographic area. 

We begin by detailing the decision variables. Let $v_{c}^{d} \in \{0,1\}$ be a boolean decision variable with value 1 if an \gls{SRE} device of technology $d\in \mathcal{D}$ is installed in \gls{cs} $c\in\mathcal{C}$ and 0 otherwise. As such, we call $v_{c}^{d}$ device installation variable. 


We can now give the FCMC formulation:
\begin{subequations}
\begin{align}
    \textbf{FCMC} \quad &\min \sum_{c \in \mathcal{C}, d \in \mathcal{D}} O^d v_c^d \label{opt:fcmc:obj} \\
    \text{s.t.} \quad 
    &\Delta_{t}^\text{BS} + \sum_{c \in \mathcal{C}, d \in \mathcal{D}} \Delta_{c,t}^d v_{c}^d \geq K 
    \quad \forall t \in \mathcal{T} \label{opt:fcmc:cov} \\
    &v_c^d \in \{0,1\} \quad \forall c \in \mathcal{C}, d \in \mathcal{D} \label{opt:fcmc:binary}
\end{align}
\end{subequations}

Here, the objective function in~(\ref{opt:fcmc:obj}) minimizes the overall cost of the network, given as the sum of all the installed \gls{SRE} devices.
%
%
Constraint~(\ref{opt:fcmc:cov}) logically expresses the coverage condition by means of the boolean link activation parameters defined in~(\ref{eq:bs_act_param}) and~(\ref{eq:ris_act_param}). 
For example, assuming $K=1$, the constraint ensures that every \gls{tp} experiences \gls{SNR} above the threshold $\Gamma$. 
If we consider $K>1$, the network layout is such that every \gls{tp} will be covered by at least $K$ distinct devices.

Although lean, this formulation is equivalent to a set-cover problem, which is proven to be NP-hard~\cite{combinatorial_optimization_book}. Nonetheless, heuristic approaches are not required, as large and realistic planning instances can be solved quite quickly, as will be shown in Sect.~\ref{sect:results}.
\subsection{Maximum Budget-Constrained Coverage (MBCC)}
FCMC produces a least-cost network layout where every \gls{tp} is guaranteed to experience an \gls{SNR} above threshold $\Gamma$. As an alternative, we also propose the MBCC formulation where the number of covered \glspl{tp} is maximized (but not guaranteed to be all), with an overall cost constrained to a budget. 

MBCC reuses the same notation used in FCMC, with the addition of a coverage boolean decision variable $\nu_{t} \in \{0,1\}$ that equals 1 if \gls{tp} $t \in \mathcal{T}$ is covered (i.e., \gls{SNR} $\geq \Gamma$) and 0 otherwise. 

\begin{subequations}
\begin{align}
    \textbf{MBCC} \quad &\max \sum_{t \in \mathcal{T}} \nu_t&\label{opt:mbcc:obj}\\
    s.t. \quad &\Delta_{t}^\text{BS} \;\;+ \sum_{c \in \mathcal{C}, d \in \mathcal{D}} \Delta_{c,t}^d v_{c}^d \geq \nu_t & \forall t \in \mathcal{T}\label{opt:mbcc:cov}\\
     &\sum_{c\in \mathcal{C}, d \in \mathcal{D}} O^d v_c^d\leq B&\label{opt:mbcc:budget}\\
     &\nu_t \in \{0,1\}\quad \forall t \in \mathcal{T}\\
    &v_c^d\in \{0,1\}\quad \forall c\in \mathcal{C}, d \in \mathcal{D}
\end{align}
\end{subequations}

As previously mentioned, the objective in~(\ref{opt:mbcc:obj}) maximizes the coverage.
Constraint~(\ref{opt:mbcc:cov}) is quite similar to~(\ref{opt:fcmc:obj}), with the difference that it allows the coverage variables $\nu_t$ to be 1 only if the corresponding \gls{tp} $t\in \mathcal{T}$ can be adequately covered by the BS or by any installed \gls{SRE} device.
Constraint~(\ref{opt:mbcc:budget}) limits the overall cost of the network to not exceed the budget $B$. 

Note that the MBCC formulation is equivalent to a maximum coverage problem, which is thus NP-hard~\cite{combinatorial_optimization_book}. However, similarly to FMCM, large instances can be solved with ease. 

\section{Results\label{sect:results}}
In this section, we present numerical results obtained using two distinct set of devices: a reduced set and a full set. The reduced set comprises well-established technologies, specifically \gls{RIS} and \gls{NCR}, which are already prototyped or deployed in practical networks with known performance characteristics. The full set includes next-generation devices, namely \gls{STAR} and \gls{3SNCR}, in addition to the reduced set. This comparison aims to evaluate whether advanced devices offer substantial performance improvements over conventional options and to assess the trade-offs between performance gains and increased costs. The analysis further underscores the potential benefits of integrating emerging technologies into future networks.

For each scenario, the optimization models are applied to eight different $400 \mathrm{m} \times 400 \mathrm{m}$ areas within the Milan map, chosen as representative of urban environments. A single \gls{BS} is placed approximately in the center of each testing area, placed on the roof of a building at a height of $10$m. For simplicity, all buildings are assumed to have a uniform height of $6,\mathrm{m}$ due to the unavailability of precise height data.
\glspl{RIS} are mounted on the exterior walls of buildings at heights of $5$, while \glspl{NCR} and \glspl{STAR} are installed on the roofs at a height of $6.5,\mathrm{m}$. Unless otherwise specified, the parameters in Table~\ref{tab:SimParam_2} are used throughout the simulations. 
\begin{table}[thb!]
    \centering
    \footnotesize
    \caption{Default simulation parameters used in Section \ref{sect:results}.}
    \begin{tabular}{l|c|c}
    \toprule
        \textbf{Parameter} &  \textbf{Symbol} & \textbf{Value(s)}\\
        \hline
        Carrier frequency & $f_0$  & $28$ GHz \\
        Bandwidth & $B$ & $200$ MHz\\
        BS Transmit power & $\sigma^2_s$ & $35$ dBm\\
        Noise power & $\sigma^2_n$ & $-82$ dBm\\
\gls{NCR} array size \cite{SRE} & $N_{p,h}\times N_{p,v}$ & $12 \times 6$ 
\\\gls{NCR} amplification gain \cite{SRE} & $\vert  g\vert^2$           & $55$ dB\\ \gls{NCR} E2E gain & $G$  & $95$ dB 
\\
\gls{NCR} noise figure                  & $NF_{ncr}$           & $8$ dB 
\\
\gls{NCR} panel separation \cite{RIS_vs_NCR_Reza}& $\alpha$  & 120 deg\\
        \gls{RIS}/\gls{STAR} elements & $M \times M$ & $100 \times 100$\\
        \gls{RIS} elements's spacing & $d_n,d_m$ & $\lambda_0/2$ m\\
        BS antenna array dimension & $N_\mathrm{t} $ & $12\times 16$ \\
        Tx elements's spacing & $d_\mathrm{Tx}$ & $\lambda_0/2$ m \\
        BS height \cite{SG_Agg_Inter} & $h_\mathrm{BS}$ & 10 m \\
        UE height & $h_\mathrm{UE}$ & 1.5 m \\
        \gls{RIS} height & $h_\mathrm{ris}$ & 5 m \\
        STAR and \gls{NCR} height & $h_\mathrm{star}$, $h_\mathrm{ncr}$ & 6.5m  \\
        Device prices in default settings & $O^\mathrm{ris}$, $O^\mathrm{ncr}$ & 1, 3  \\
Blocker height  \cite{DynamicBlockage}                                                    & $h_{B}$         & 1.7 m
\\ 
Blocker density \cite{DynamicBlockage}                                                     & $\lambda_{B}$           & $4\times 10^{-3}$ m$^{-2}$\\
Blocker velocity \cite{DynamicBlockage}                           & $V$           & 15 m/s
\\
Blockage duration \cite{DynamicBlockage}                           & $\mu^{-1}$            & 5 s
\\ 
        \bottomrule
    \end{tabular}
    \label{tab:SimParam_1}
\end{table}

\begin{figure*}[!htb]
\centering
\subfloat[][\small Reduced set of devices]{\includegraphics[width=0.45\textwidth]{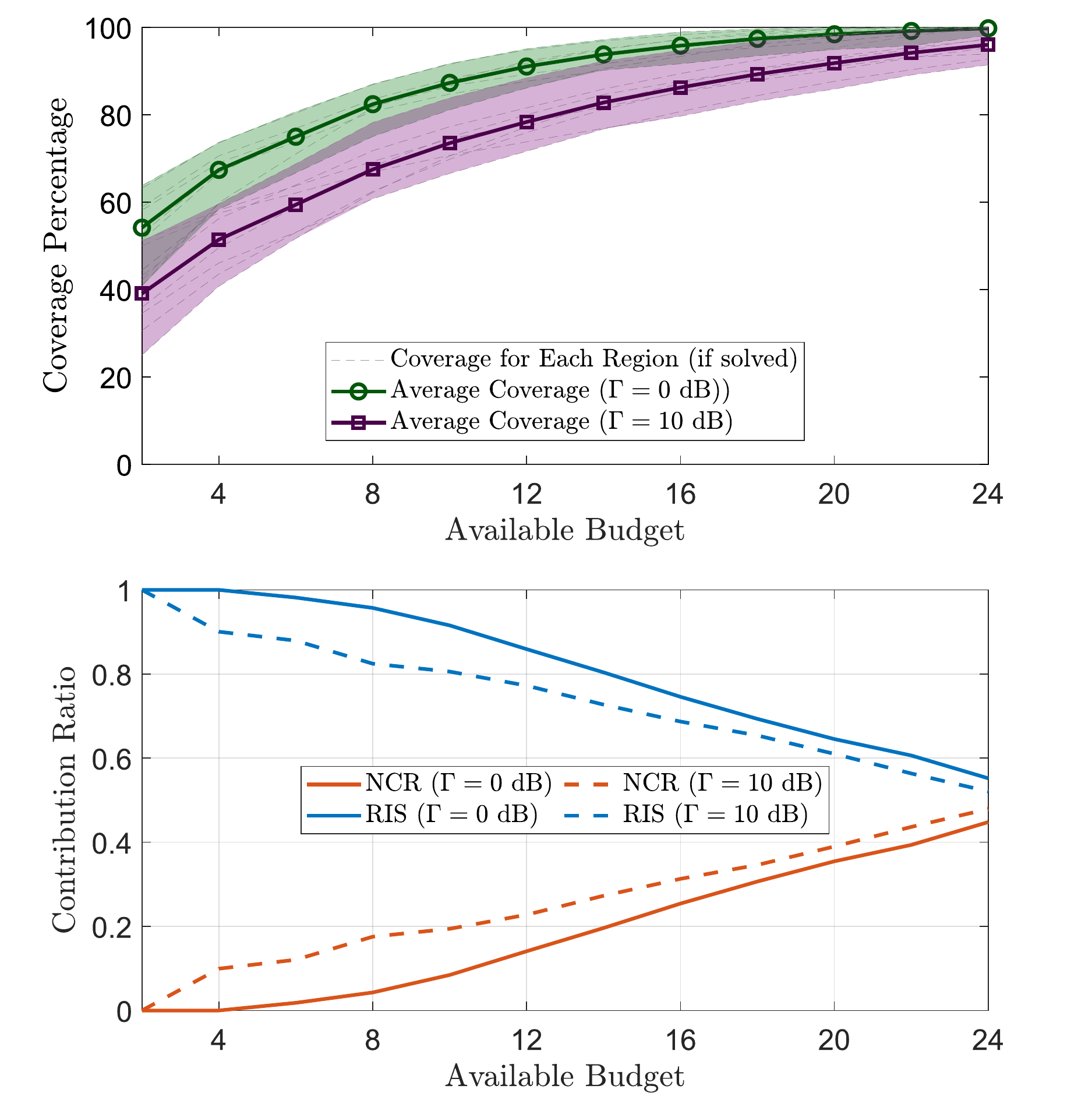}}\hspace{0.7cm}
\subfloat[][\small Full set of devices]{\includegraphics[width=0.45\textwidth]{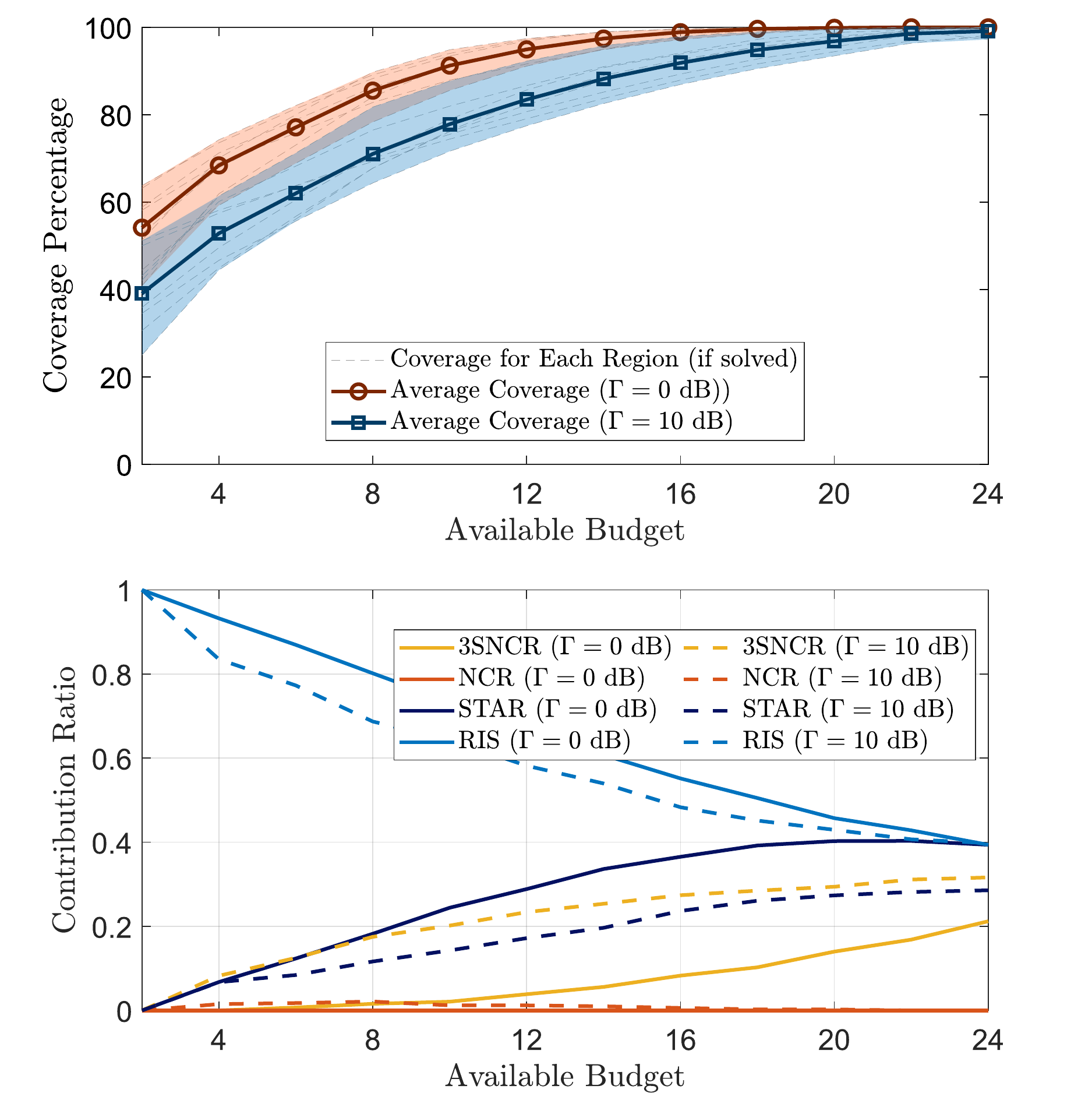}}

\caption{Average Coverage probability (achieved with MBCC optimization model) vs the total available budget. The used configurations are the default as in Table \ref{tab:SimParam_1}. }\label{fig:Abudget}
\end{figure*} 
\subsection{Results with \gls{MBCC} Optimization Model}\label{sec:MBCC_examples}
This subsection presents numerical results using the \gls{MBCC} optimization model. 
Figures~\ref{fig:Abudget}-(a) and \ref{fig:Abudget}-(b) illustrate the maximum coverage percentage as a function of the available budget for the reduced and full sets of devices, respectively. The gray dashed lines represent the results for eight simulated urban regions, while the shaded areas indicate the upper and lower coverage bounds across these regions. The solid lines with markers represent the average values. In each figure, the top panel shows the maximum coverage percentage and the bottom panel shows the contribution ratio, which indicates the relative proportion of each device type among all devices deployed \gls{SRE}. This representation highlights device behavior and contribution trends under varying budget conditions.

\begin{figure*}[!t]
\centering
\subfloat[][\small Reduced set of devices]
{\includegraphics[width=0.45\textwidth]{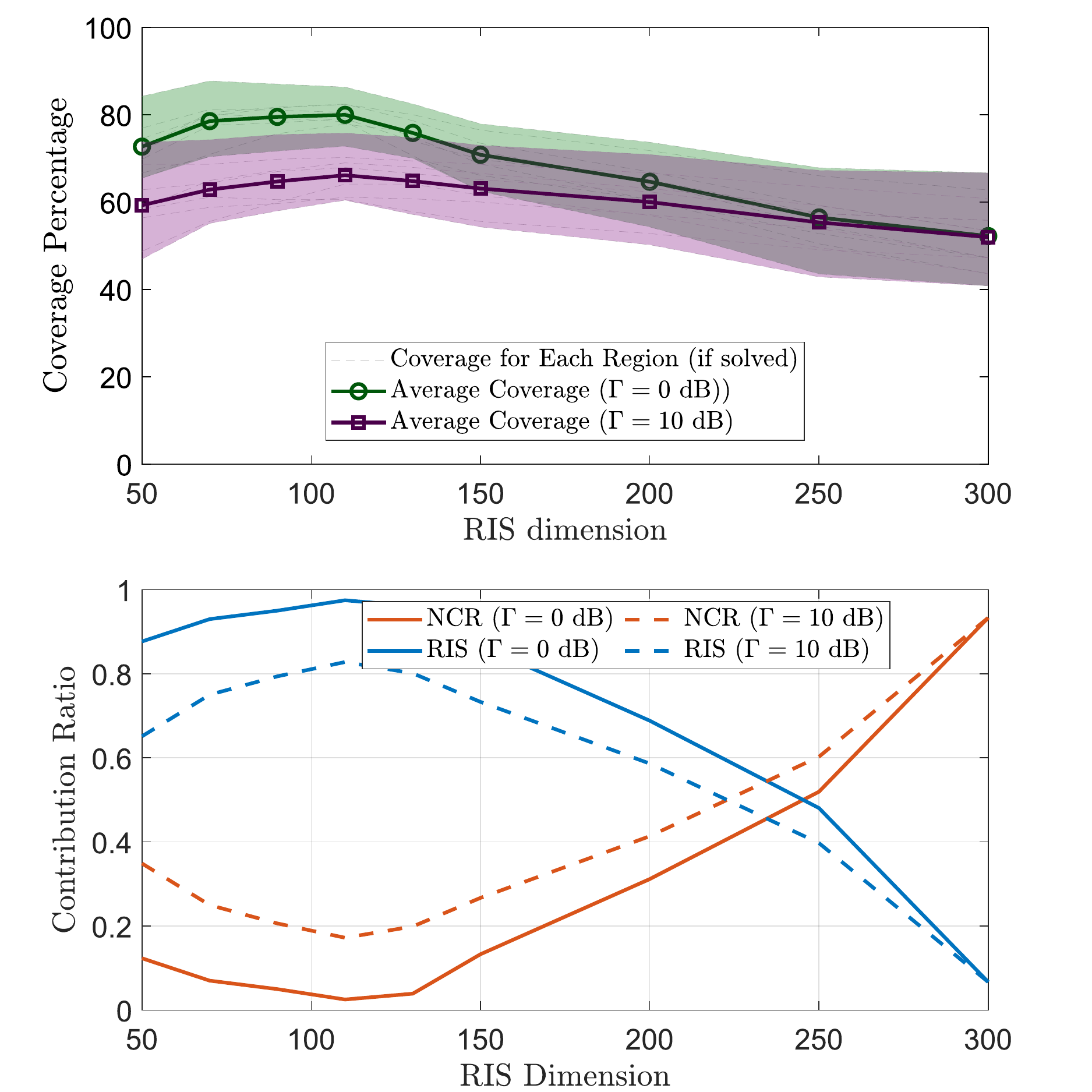}}\hspace{0.7cm}
\subfloat[][\small Full set of devices]
{\includegraphics[width=0.45\textwidth]{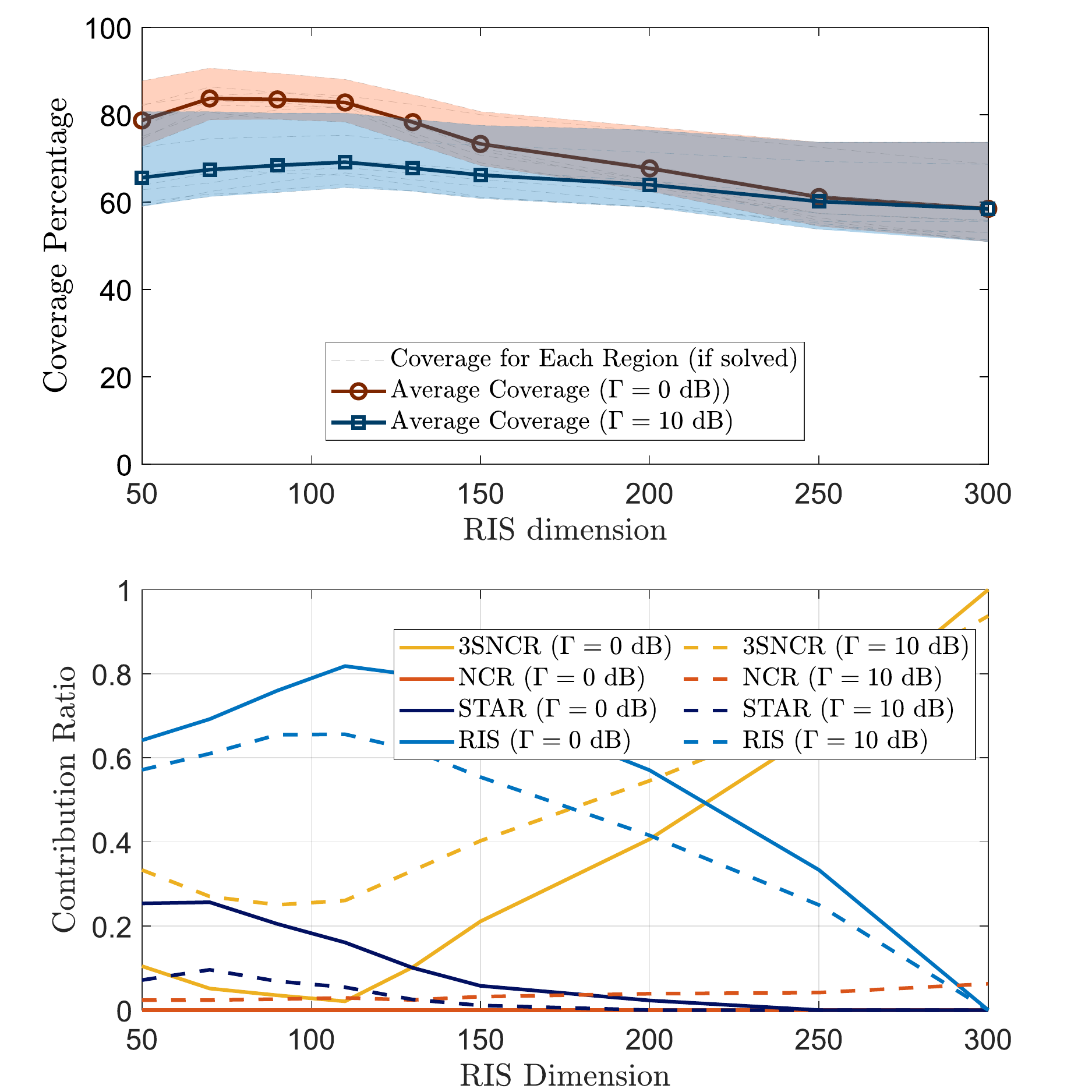}}
 \caption{Average Coverage probability vs \gls{RIS} dimension, where the total available budget is constrained to $B=8$ units. }\label{fig:opt1_ris_size}
\end{figure*}

\begin{figure*}[!htb]
\centering
\subfloat[][\small Reduced set of devices]
{\includegraphics[width=0.45\textwidth]{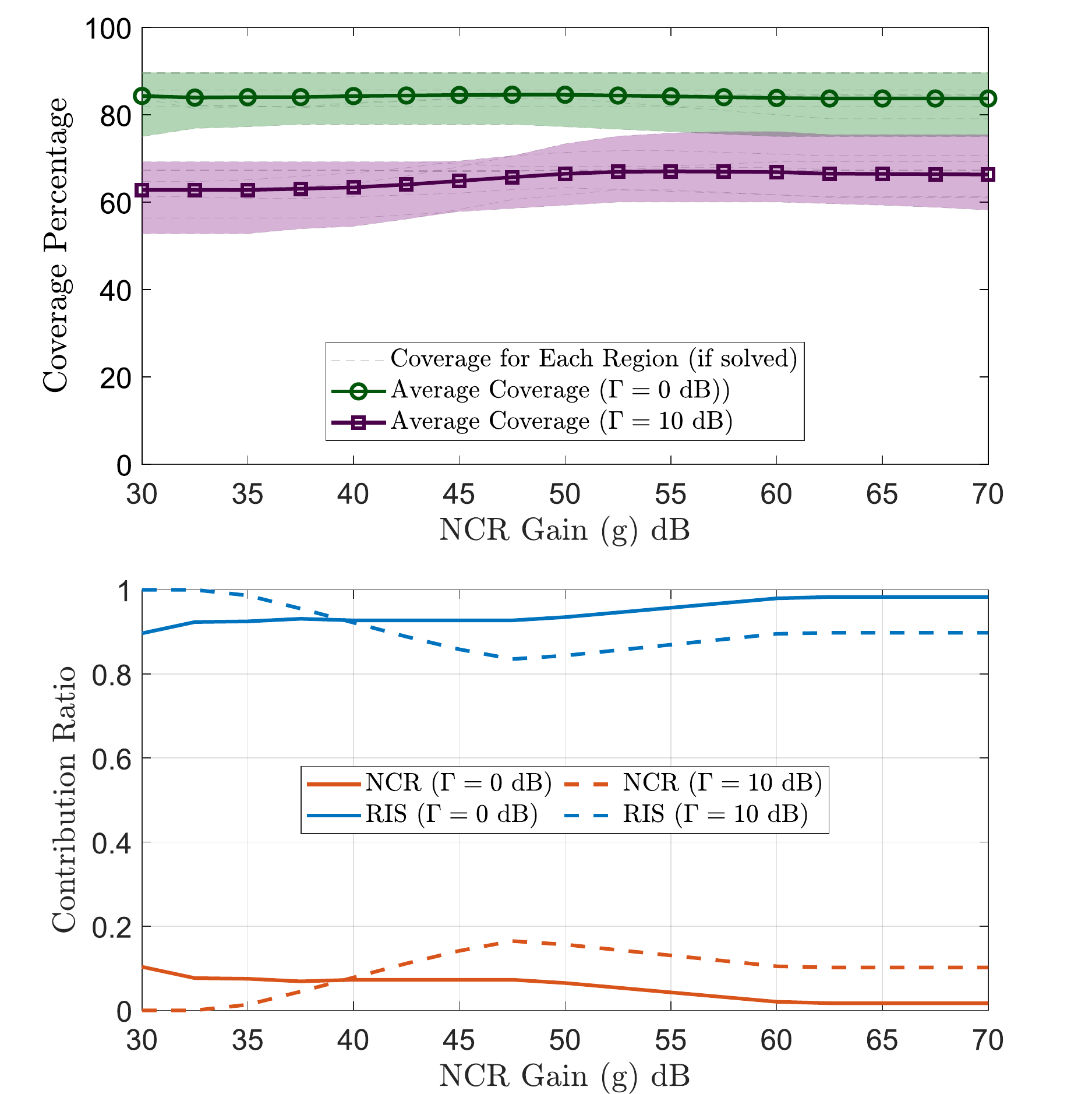}}\hspace{0.7cm}
\subfloat[][\small Full set of devices]
{\includegraphics[width=0.45\textwidth]{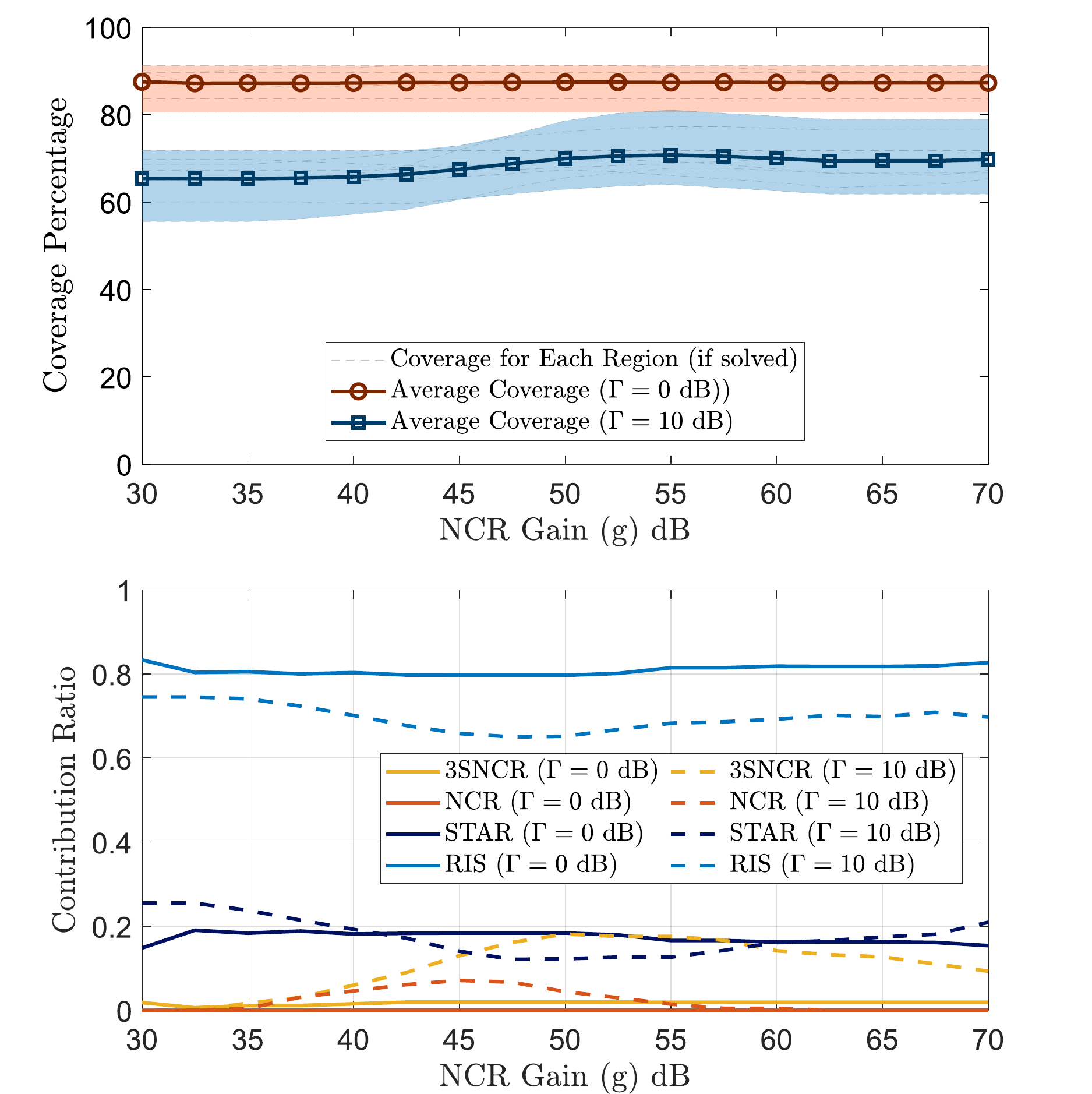}}
 \caption{Average Coverage probability (achieved with MBCC optimization model) vs \gls{NCR} amplification gain, where the total available budget is constrained to $B=8$ units. Configurations according to table \ref{tab:SimParam_1}. }\label{fig:opt1_ncr_gain}
\end{figure*}

As shown in Fig.~\ref{fig:Abudget}-(a), the reduced device set demonstrates a clear trend: the contribution of \gls{NCR}s increases with higher budgets, while the contribution of \gls{RIS}s decreases. This behavior occurs because \gls{NCR}s, despite their higher cost, effectively serve blind spots behind buildings and distant users due to their amplification gain, overcoming the reflective limitations of \gls{RIS}s. With increasing budgets, the model prioritizes \gls{NCR}s, a pattern observed for both $[\Gamma]_{\textrm{dB}} = 0$ and $[\Gamma]_{\textrm{dB}} = 10$.

Figure~\ref{fig:Abudget}-(b) highlights similar trends for the full set of devices, with notable differences. The model favors \gls{3SNCR}s over \gls{NCR}s due to their wider field of view without additional cost. As the budget increases, the contribution of \gls{RIS} s decreases, but instead of relying solely on \gls{3SNCR}s, \gls{STAR} devices play an increasing role. This is because \gls{STAR}s, when placed on rooftops, can serve users behind buildings using refraction, a function that would otherwise require \gls{NCR}s in the reduced set. With higher budgets, the contribution of \gls{STAR} s continues to increase steadily.

In subsequent analyses, device configurations and costs are scaled to explore their impact. For example, Fig.~\ref{fig:opt1_ris_size} shows the percentage of coverage versus \gls{RIS} dimensions, assuming a fixed total budget of $B = 8$ units. This budget can ensure $80\%$ coverage for $\Gamma = 0$~dB, as shown in Fig.~\ref{fig:Abudget}. As the dimensions of \gls{RIS} increase from $M = 50\times 50$ to $M = 300\times 300$, the coverage percentage initially improves, peaks, and then declines. This behavior suggests an optimal \gls{RIS} size for coverage.

For example, with the reduced set in Fig.~\ref{fig:opt1_ris_size}-(a) and $[\Gamma]_{\textrm{dB}} = 0$~dB, increasing \gls{RIS} dimensions up to $M = 120 \times 120$ enhances coverage due to better channel gain. Beyond this point, the increasing cost of larger \gls{RIS}s outweighs their performance benefits, as the additional gain is unnecessary to meet the \gls{SNR} threshold. A similar trend is evident in the contribution ratio: initially, \gls{RIS} contributions increase with size but decrease beyond $M = 120\times 120$, with \gls{NCR}s or \gls{3SNCR}s taking over. In Fig.~\ref{fig:opt1_ris_size}-(b), with the full set of devices, the contributions \gls{STAR} also decrease as the number of cells in the metasurface increases, indicating that the metasurfaces are effective when their cost is limited.
\begin{figure*}[!htb]
\centering
\subfloat[][\small Reduced set of devices, $K=1$]
{\includegraphics[width=0.45\textwidth]{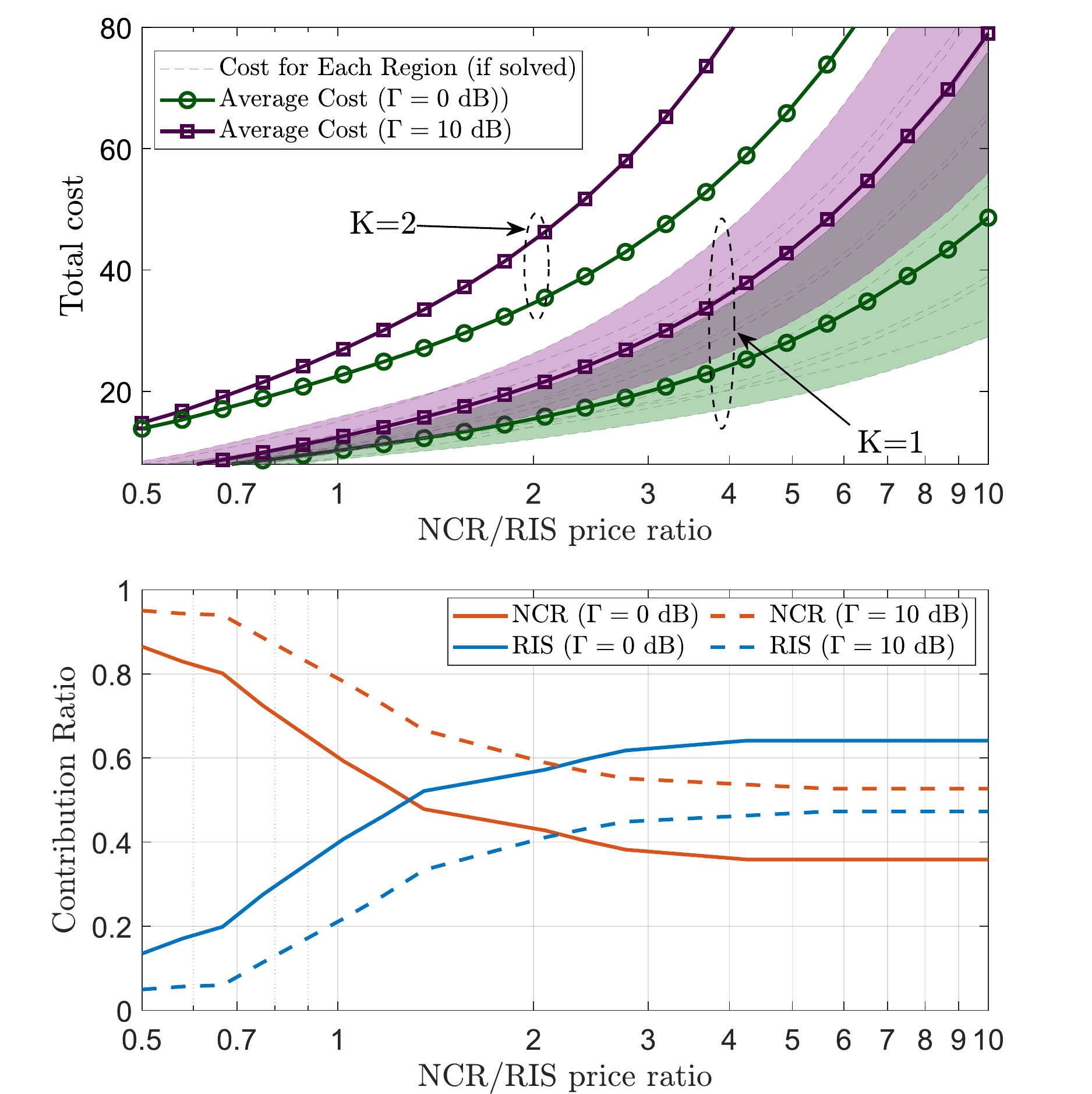}}\hspace{0.7cm}
\subfloat[][\small Full set of devices, $K=1$]
{\includegraphics[width=0.45\textwidth]{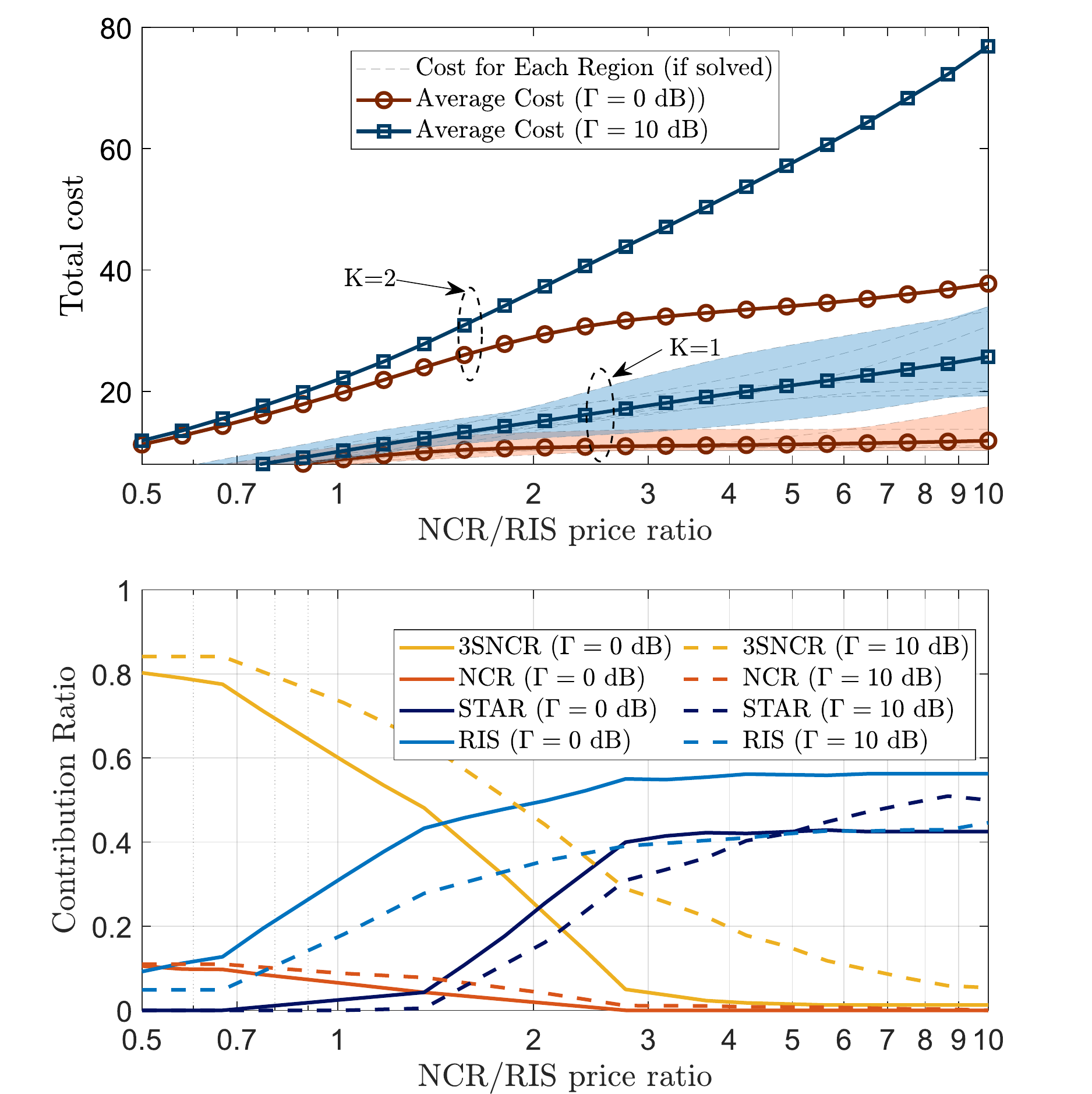}}
 \caption{Average total cost (achieved with FCMC optimization model) vs NCR/\gls{RIS} price ratio with FCMC optimization model, for fixed settings of \gls{RIS}/\gls{NCR} (see table \ref{tab:SimParam_1}).  }\label{fig:opt2_ratio}
\end{figure*}

Figure~\ref{fig:opt1_ncr_gain} examines the impact of \gls{NCR} (or \gls{3SNCR}) amplification gain on coverage. Here, the configurations of \gls{RIS} are fixed to the default values of Table~\ref{tab:SimParam_1}, and the total budget remains $B = 8$ units. For $[\Gamma]_{\textrm{dB}} = 0$, increasing \gls{NCR} gain initially reduces their contribution as their cost increases, causing the model to favor \gls{RIS} or \gls{STAR} devices, which still satisfy the \gls{SNR} requirement.

For $[\Gamma]_{\textrm{dB}} = 10$, the trend differs. Initially, higher \gls{NCR} gain improves both the coverage percentage and \gls{NCR} contribution. However, beyond amplification gains of $[g]_{\textrm{dB}} = 47$ for the reduced set (Fig.~\ref{fig:opt1_ncr_gain}-(a)) and $[g]_{\textrm{dB}} = 52$ for the full set (Fig.~\ref{fig:opt1_ncr_gain}-(b)), \gls{NCR} contributions decline as their costs outweigh their benefits.
\begin{figure*}[!htb]
\centering
\subfloat[][\small Reduced set of devices ]
{\includegraphics[width=0.45\textwidth]{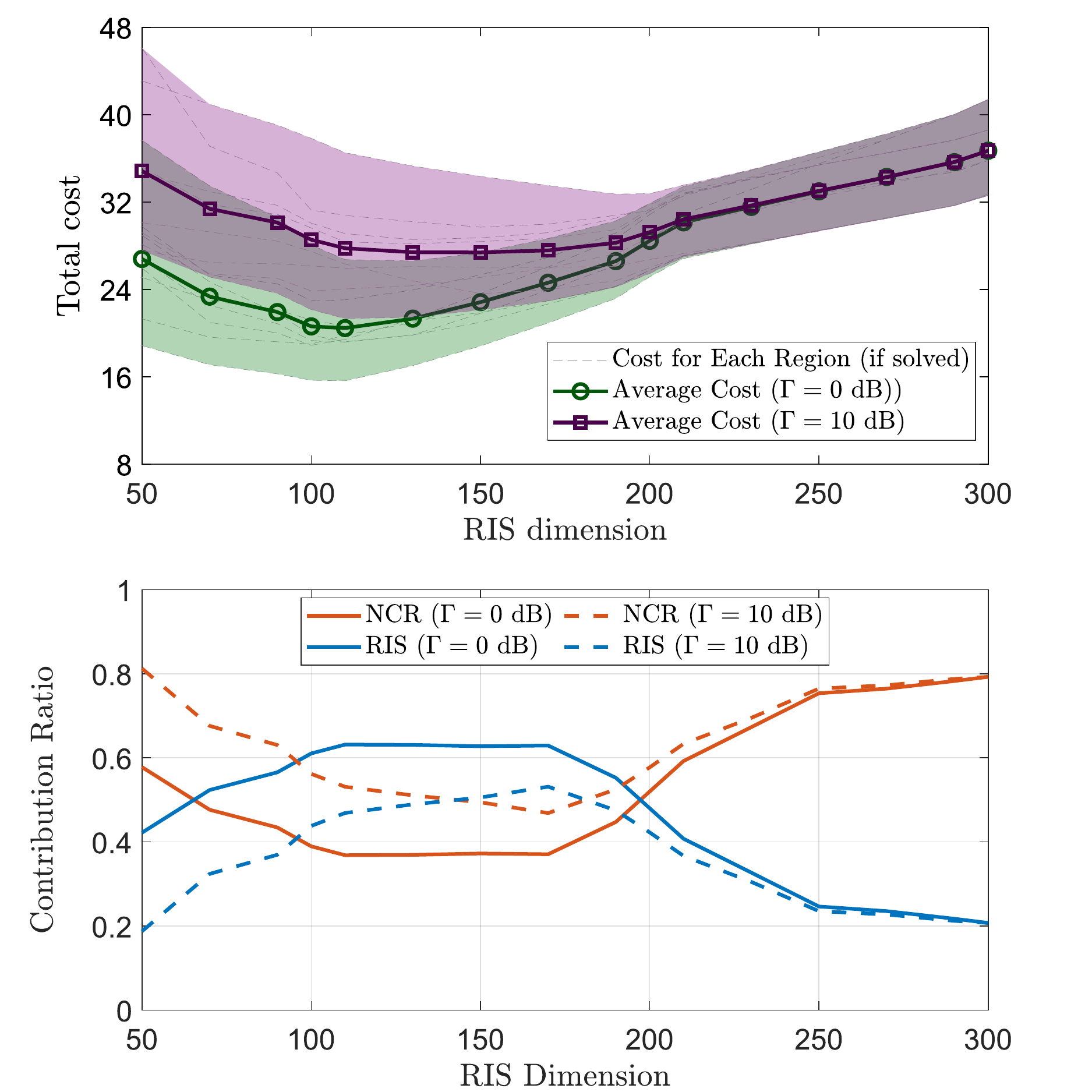}}\hspace{0.7cm}
\subfloat[][\small Full set of devices Devices]
{\includegraphics[width=0.45\textwidth]{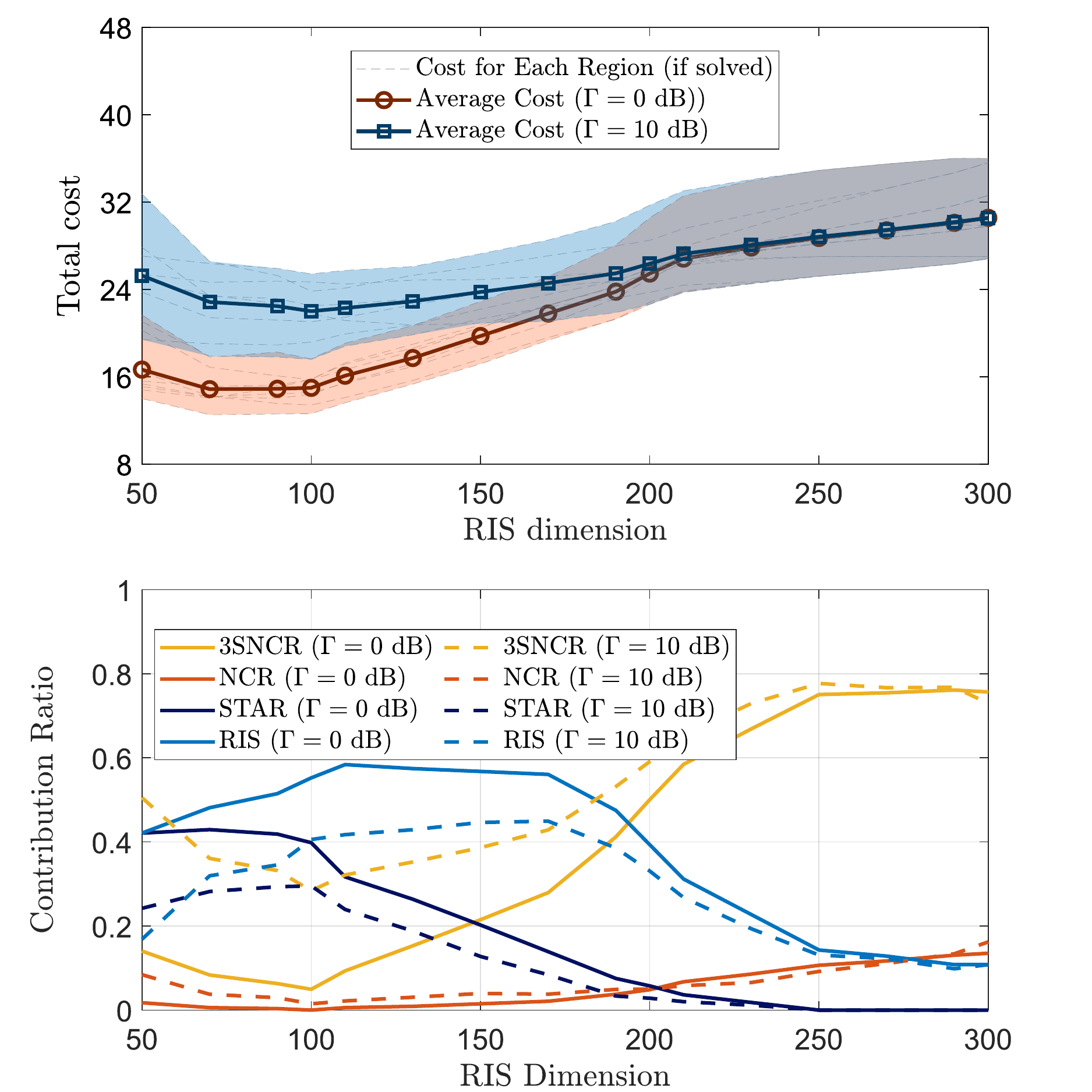}}
 \caption{Average total cost (achieved with FCMC optimization model) vs \gls{RIS} side size with, where the configurations and price of the \gls{NCR} is the default according to table \ref{tab:SimParam_1} and the cost of the \gls{RIS} scales according to table \ref{tab:SimParam_2}. }\label{fig:opt2_RIS_dim}
\end{figure*}

\begin{figure*}[!htb]
\centering
\subfloat[][\small Reduced set of devices]{\includegraphics[width=0.45\textwidth]{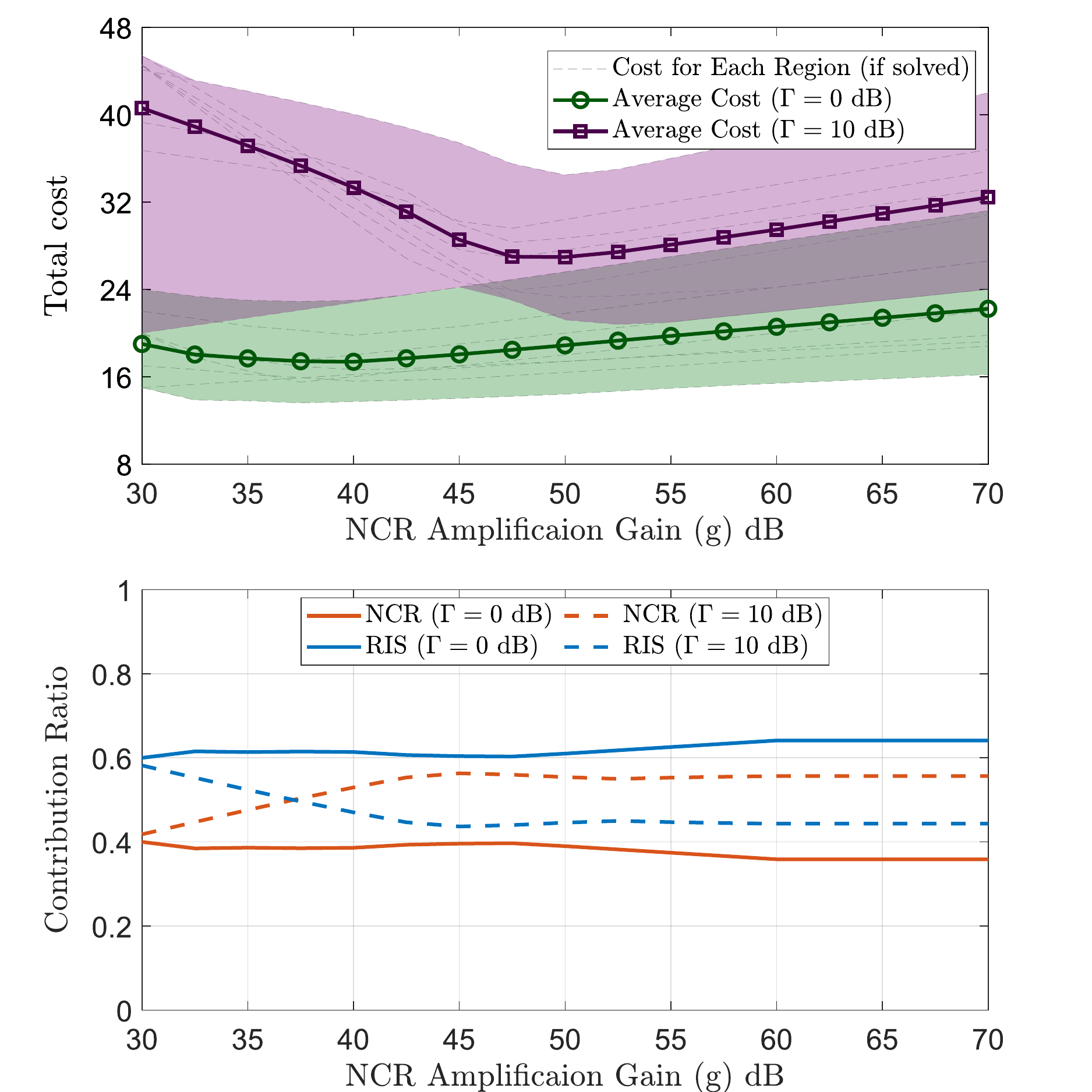}}\hspace{0.7cm}
\subfloat[][\small Full set of devices]{\includegraphics[width=0.45\textwidth]{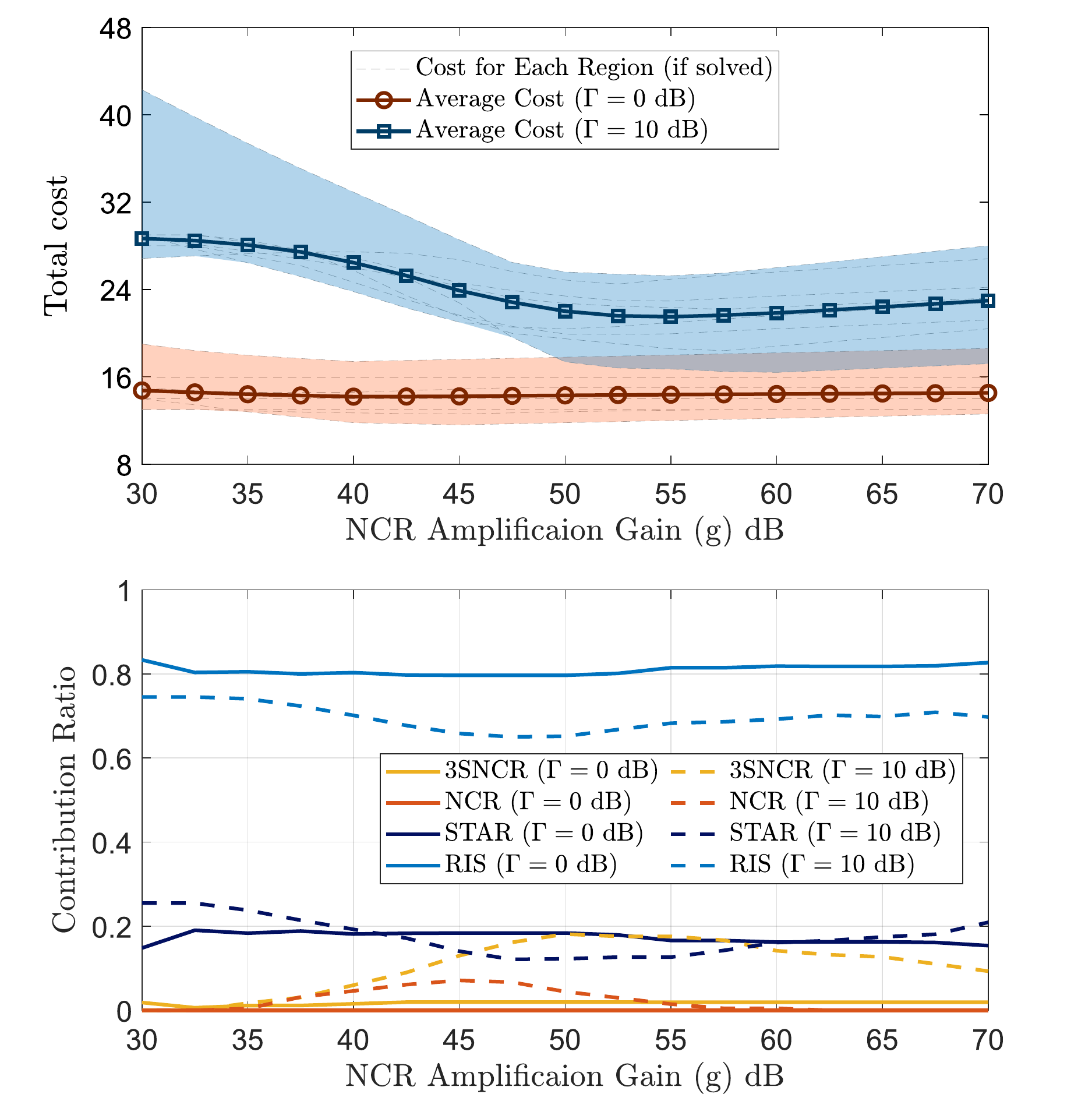}}
\caption{Average total cost (achieved with FCMC optimization model) vs \gls{NCR} end-to-end gain (G), where the \gls{RIS} has default configurations as table \ref{tab:SimParam_1}, and the cost of \gls{NCR} scales according to table \ref{tab:SimParam_2}.}\label{fig:opt2_ncr_gain}
\end{figure*}

\begin{figure*}[t!]
    \centering
    \subfloat[]{\includegraphics[width=0.33\textwidth]{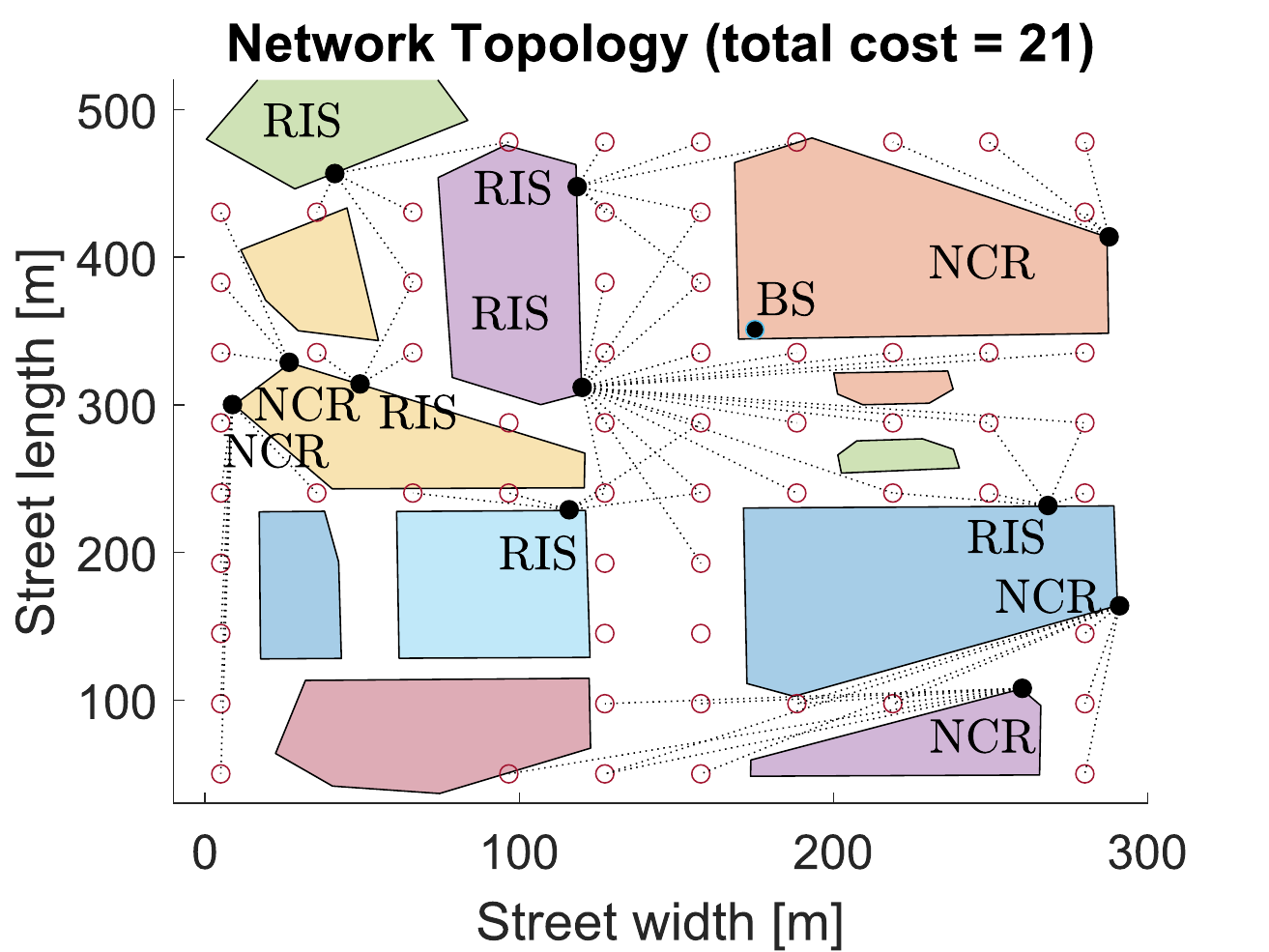}} 
    \subfloat[]{\includegraphics[width=0.33\textwidth]{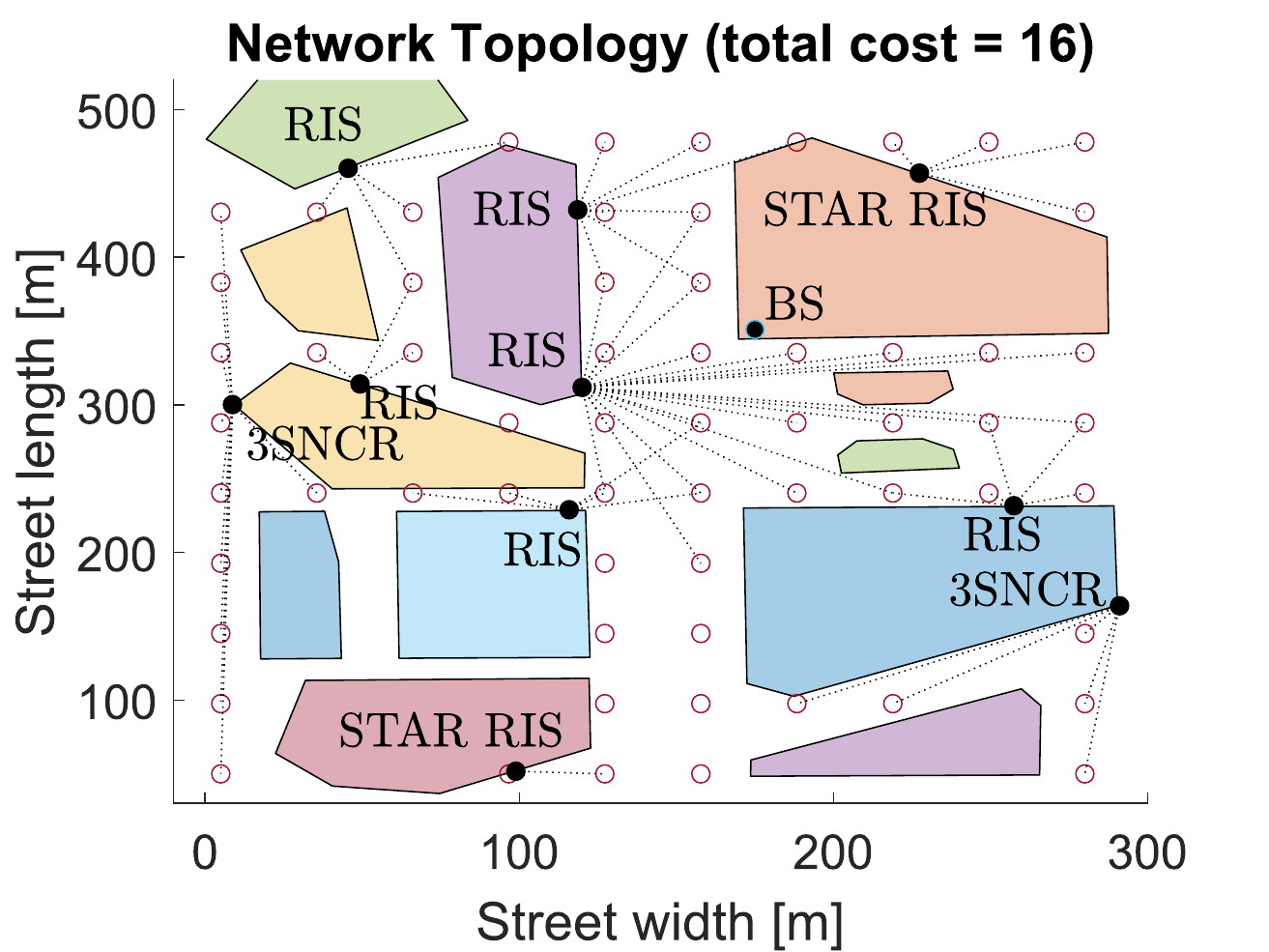}}
    \subfloat[]{\includegraphics[width=0.33\textwidth]{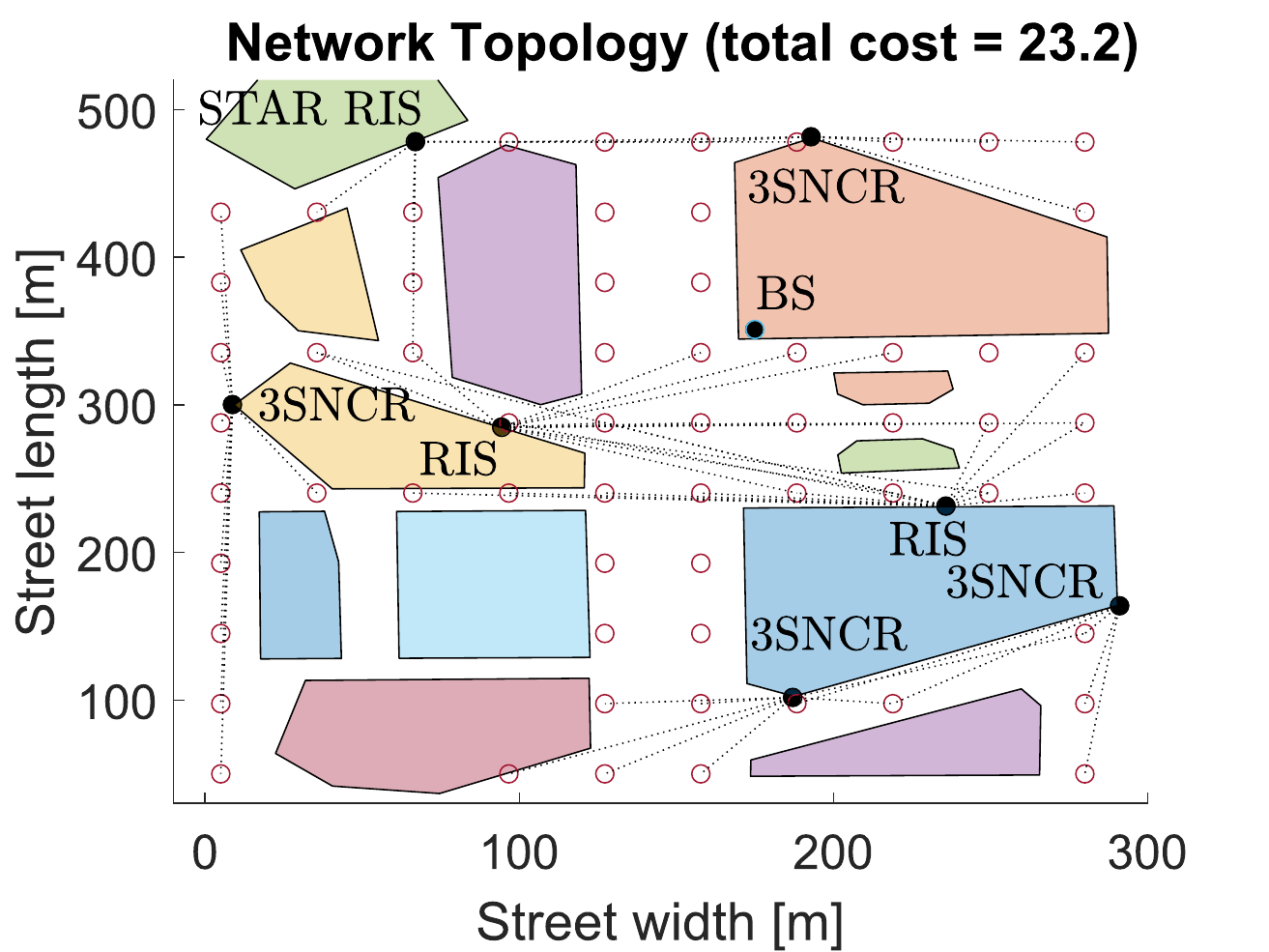}}
    \caption{Network topology, achieved by \gls{FCMC} optimization model, when using (a) the reduced set of devices with default configurations and prices as in table \ref{tab:SimParam_1} (b) full set of devices with default configurations and prices as in table \ref{tab:SimParam_1}, (c) full set of devices with RIS dimension $M=200\times 200$ and RIS cost scaling accoding to Table \ref{tab:SimParam_2}. The \gls{NCR} costs and configurations are always assumed default according to table \ref{tab:SimParam_1}. }\label{fig:map_planning}
\end{figure*}

\subsection{Results with \gls{FCMC} Model}\label{sec:FCMC_examples}
In this set of examples, we analyze the performance of the \gls{FCMC} optimization model, which minimizes the cost required to achieve full coverage of all \gls{tp}s. Figure~\ref{fig:opt2_ratio} shows the total cost as a function of the \gls{NCR}-to-\gls{RIS} price ratio for both reduced and full sets of devices. As the price ratio increases, the total cost increases exponentially, particularly with the reduced set of devices (Fig.~\ref{fig:opt2_ratio}-(a)). This rapid growth occurs because, in this model, all \gls{tp}s must be covered. \gls{tp}s far from \gls{BS} or blocked by buildings can only be served by \gls{NCR}s in the reduced set, regardless of their cost.

In contrast, the full set of devices (Fig.~\ref{fig:opt2_ratio}-(b)) mitigates this issue by replacing costly \gls{NCR}s/\gls{3SNCR}s with \gls{STAR}s, which can transmit toward \gls{tp}s behind buildings when placed on the roofs. For $[\Gamma]_{\textrm{dB}} = 0$, the contribution of \gls{RIS}s surpasses \gls{NCR}s when the cost ratio exceeds $\ O ^{ncr}/\ O ^{ris} > 1.3$. However, at $[\Gamma]_{\textrm{dB}} = 10$, the contribution of \gls{NCR} remains significant even at higher cost ratios, exceeding $\ O ^{ncr}/\ O ^{ris} > 3$, because they achieve higher \gls{SNR} requirements. With the full set, \gls{RIS}s surpass \gls{3SNCR}s when $\ O ^{ncr}/\ O ^{ris} > 1.4$. For $\ O ^{ncr}/\ O ^{ris} > 3$, \gls{NCR}s/\gls{3SNCR}s are entirely replaced by \gls{RIS}s and \gls{STAR}s.

Furthermore, Fig.~\ref{fig:opt2_ratio} includes results for $K = 2$, where at least two devices serve each \gls{tp}. As expected, the total cost for $K = 2$ is significantly higher than for $K = 1$. However, the full set of devices helps moderate this increase compared to the reduced set. By setting \( K = 2 \), the planning becomes more robust against dynamic blockages that may occur during the online operation of the network. 

Similarly to the \gls{MBCC} model (Section~\ref{sec:MBCC_examples}), we scale the price of the devices based on their configurations (\gls{RIS} dimension or \gls{NCR} gain). Initially, the \gls{NCR} gain is fixed according to Table~\ref{tab:SimParam_1}, and the \gls{RIS} configurations are varied. Figure~\ref{fig:opt2_RIS_dim} shows the total cost as a function of the \gls{RIS} dimension $\sqrt{M}$.
For the reduced set of devices (Fig.~\ref{fig:opt2_RIS_dim}-(a)), increasing the \gls{RIS} size from $M = 50 \times 50$ to $M = 120 \times 120$ for $[\Gamma]_{\textrm{B}} = 0$ and to $M = 150 \times 150$ for $[\Gamma]_{\textrm{dB}} = 10$, reduces the total cost by approximately 20\%. However, beyond these dimensions, the total cost increases due to the increased expense of larger \gls{RIS} size.
For the full set of devices (Fig.~\ref{fig:opt2_RIS_dim}-(b)), the optimal \gls{RIS} (or \gls{STAR}) dimension that minimizes the total cost is approximately $M = 100 \times 100$. Beyond this size, the total cost increases, suggesting that fewer metasurface cells are needed to achieve optimal coverage when advanced devices such as \gls{STAR} are considered. Remarkably, the total cost of the full set of devices is consistently lower than with the reduced set, and this is due to the greater flexibility in optimizing device deployment with the full set.
For certain dimensions, such as $M = 50 \times 50$, the total cost with the full set of devices (Fig.~\ref{fig:opt2_RIS_dim}-(b)) is 40\% lower than with the reduced set (Fig.~\ref{fig:opt2_RIS_dim}-(a)). These findings align with the results of the \gls{MBCC} model, confirming the optimal \gls{RIS} dimensions for cost-effective coverage under a fixed budget.

Fig.~\ref{fig:opt2_ncr_gain} examines the impact of \gls{NCR} amplification gain on total cost, with \gls{RIS} dimensions fixed to default values. For the reduced set (Fig.~\ref{fig:opt2_ncr_gain}-(a)), the minimum cost occurs at $[g]_{\textrm{dB}} = 38$ for $[\Gamma]_{\textrm{dB}} = 0$ and $[g]_{\textrm{dB}} = 48$ for $[\Gamma]_{\textrm{dB}} = 10$. At lower gains, additional \gls{NCR}s are required to meet \gls{SNR} requirements, increasing the cost. In contrast, excessively high gains add unnecessary costs without improving coverage.

The full set of devices (Fig.~\ref{fig:opt2_ncr_gain}-(b)) exhibits a similar trend, but the total cost is consistently lower due to the flexibility provided by advanced devices. For $[\Gamma]_{\textrm{dB}} = 10$, the cost reduction can reach 30\% for certain \gls{NCR} gain values. As the gain increases, \gls{NCR} (or \gls{3SNCR}) contributions initially rise but saturate beyond $[g]_{\textrm{dB}} = 45-50$, with additional gain increases only raising costs without affecting contributions.

Finally, Fig.~\ref{fig:map_planning} illustrates the network topology for Piazza Piola in Milan, planned using the \gls{FCMC} optimization model in different scenarios. The \gls{SNR} threshold is set to $[\Gamma]_{\textrm{dB}} = 10$, and the configurations and prices for the \gls{NCR} (or \gls{3SNCR}) devices are fixed according to Table~\ref{tab:SimParam_1}.

In Fig.~\ref{fig:map_planning}-(a), only the reduced set of devices is used with default \gls{RIS} configurations. In Fig.~\ref{fig:map_planning}-(b), the full set of devices is employed, again with default configurations for \gls{RIS} and \gls{STAR}. The results show a decrease in total cost to 16 units, compared to 21 units with the reduced set. This reduction is achieved by replacing two \gls{NCR}s with one \gls{3SNCR} (as their prices are equivalent) and substituting two additional \gls{NCR}s with one \gls{STAR}.

Figure~\ref{fig:map_planning}-(c) examines the effect of increasing the dimensions of \gls{RIS}/\gls{STAR} devices. In this case, the total cost increases due to the higher expense of larger \gls{RIS}/\gls{STAR} configurations. As a result, the model shifts to deploy more \gls{3SNCR}s compared to Fig.~\ref{fig:map_planning}-(b), despite the relatively high cost of \gls{3SNCR}s. Furthermore, it can be observed that the lengths of the \gls{RIS}/\gls{STAR} links are longer in this scenario, as their larger size compensates for increased path loss.

\section{Conclusion \label{sect:conclusion}}
In this paper, we develop a deployment optimization framework for \gls{HSRE}, focusing on realistic urban scenarios. Our study explored both foundational and advanced \gls{SRE} devices, specifically \gls{RIS}, \gls{STAR}, \gls{NCR}, and \gls{3SNCR}. Two optimization models were proposed: one aimed at minimizing the overall cost of complete coverage and another focused on maximizing coverage within a fixed budget.

The numerical results revealed that expanding the selection of the devices to include next-generation technologies such as \gls{STAR} and \gls{3SNCR} provided notable cost efficiency and improved coverage performance. In some scenarios, the integration of these advanced devices achieved up to 40\% cost savings compared to the use of traditional devices alone. Moreover, the analysis demonstrated that the optimal choice of devices depends on deployment specifics. RIS is advantageous for cost-effective, broad coverage, while NCR and 3SNCR are essential for boosting signal strength in challenging areas. This flexible planning framework underscores the potential advantages of incorporating emerging technologies into future networks.

\bibliographystyle{IEEEtran}
\bibliography{main.bib}
\end{document}